\documentclass[14pt,a4paper]{article}

\usepackage{fullpage}
\usepackage{graphicx}
\usepackage{multirow}
\usepackage{tikz}
\usetikzlibrary{calc}
\usepackage{xcolor}
\usepackage{amssymb}
\usepackage{amsmath}
\usepackage{tikz}
\usetikzlibrary{calc}
\usepackage{graphicx}

\begin{document}

{\centering \Large{\textbf{Kaon and strangeonium spectrum in Regge phenomenology}}}

\begin{center}
	 Juhi Oudichhya$^{1}$, Keval Gandhi$^{2}$ and Ajay Kumar Rai$^{1}$ \\
	 $^{1}$Department of Physics, Sardar Vallabhbhai National Institute of Technology, Surat, Gujarat-395007, India. \\
	 $^{2}$Department of Computer Sciences and Engineering,
	 	Institute of Advanced Research, Gandhinagar, Gujarat 382426, India.
\end{center}
\vspace{1cm}
\textbf{\large Abstract}
	In the present work, the mass-spectra of the light mesons, the kaons ($u\overline{s}$) and strangeonium ($s\overline{s}$) is systematically studied within the framework of Regge phenomenology.
Several relations between Regge slope, intercept, and meson masses are extracted with the assumption of linear Regge trajectories. Using these relations the ground state masses ($1^{1}S_{0}$ and $1^{3}S_{1}$) of the pure $s\overline{s}$ states are evaluated. Further, the Regge slopes are extracted for kaons and strangeonium to obtain the orbitally excited state masses in the ($J,M^{2}$) plane. Similarly, the values of Regge parameters are calculated in the ($n,M^{2}$) plane for each Regge trajectory and obtain the radially excited state masses of mesons lying on that Regge trajectory.  We compared our obtained spectrum with the experimental observations where available and with the predictions of other theoretical approaches. Here, we predict the possible quantum numbers of several recently observed experimental states, which still require further verification, and also evaluate the higher orbital and radial excited states that may be detected in the near future. We expect our predicted results could provide valuable information for future experimental searches for missing excited kaons and strangeonium mesons.


 \section{Introduction}

As a result of significant experimental progress in recent years, many new light hadronic states have been observed, and studying the properties of these states has been of great interest. The quark model holds that mesons are made up of quark and antiquark ($q\overline{q}$) while baryons are composed of three quarks ($qqq$).
In addition to these, many exotic states have also been detected experimentally, comprises of more than three quarks. Experimental evidence of new hidden charm tetraquark states, the $P_{c}$ pentaquark states \cite{exotic1,exotic2}, the hexaquark candidates \cite{exotic3} has been observed in the past two decades. In a recent study, the Heptaquark states composed of seven quarks, having two charmed mesons and one nucleon is investigated \cite{exotic4}. 
In the present work, we concentrate on the light strange and strangeonium states. 

Several strange mesons have been discovered and observed over the course of many years by various experimental facilities, such as BaBar \cite{BaBar2008}, Belle \cite{Belle2010}, LHCb \cite{LHCb2014,LHCb2017,LHCb2021}, BESIII \cite{BESIII2019,BESIII2020}  etc.
Recently, $e^{+}e^{-}$ collision experiments in BESIII and BaBaR provided data over 2.0-2.2 GeV with excellent precision, which  helped to study the excitations of light mesons \cite{Exp1,Exp2,Exp3,Exp4}. There are  ideas and plans for future measurements of strange meson at other experimental facilities, including LHCb, Belle II, and BESIII Collaboration employing either $\tau$ or $D-$ meson decays and also the upcoming experimental facilities $\overline{P}$ANDA seeking for these strange mesons \cite{PANDA}.
In addition, J-PARC will soon construct a new kaon beam line \cite{K. Aoki2021}.
In the most recent Particle Data Group (PDG) review \cite{PDG}, fourteen kaon mesons with their corresponding spin and parity are firmly known, but eleven states still need to be confirmed, with two of them; $K(1630)$ and $K(3100)$ still awaiting the determination of their spin-parity quantum numbers. In the present work, we obatined the mass spectra of kaon and tried to assign the possible spin-parity to the observed states.

Other than strange mesons, an interesting quarkonium state named  strangeonium ($s\overline{s}$) states predicted in the quark model, which are bound states of the $s$-quark and anti-$s$ quark lie between the light up, down quarks and heavy charm, bottom quarks. The study of strangeonium states is related to non-$q\overline{q}$ states (glueballs, hybrids, and tetraquarks, etc.) with the same quantum numbers as conventional $q\overline{q}$ systems. For experimental confirmation of a non-$q\overline{q}$ states, one must have a thorough understanding of the conventional $q\overline{q}$ states.
There are very few strangeonium resonances listed in the PDG \cite{PDG} that have been experimentally well confirmed and are widely accepted as pure $s\overline{s}$  states.
Except from certain low-lying $1P$ and $1D$ states, many excited strangeonium states  have yet to be detected in near future. To study these states, the BESIII experiment provides a strong platform by not only confirming the previously observed $s\overline{s}$ states but also discovering some new resonances by the decays of $J/\psi$ and $\psi(2S)$ \cite{P. L. Liu2015,C. Z. Yuan2019,M. Ablikim2020,ssbar1,ssbar2,ssbar3}.
For instance, in 2016 several states around 2.0-2.4 GeV were detected at BESIII, the $f_{2}(2010)$ resonance mentioned in PDG \cite{PDG} was confirmed in this process. Another state $X(2500)$ observed with mass $M$ = 2470$^{+15+101}_{-19-23}$ MeV still needs more confirmation \cite{ssbar3}. 

In 2018, the BESIII Collaboration detected several isoscalar mesonic states, out of which the $f_{0}(2330)$ state with $J^{PC}$ = $0^{++}$ might be a good candidate of strangeonium state \cite{ssbar2}. Experimentally, this resonance still needs more confirmation. The LASS Collaboration observed a narrow resonance $f_{4}(2210)$ with $J^{PC}$ = $4^{++}$ having mass 2209$^{+17}_{-15}$ MeV \cite{LASS1988}. Later on, the results obtained by the MARK-III Collaboration \cite{MARKIII} and WA67 (CERN SPS) \cite{PSLBooth1986} matches with the observation of \cite{LASS1988}. This resonance is still not confirmed and needs further verification.
Recently, in 2019 the evidence of new resonances $X(2060)$ having mass M = (2062.8$\pm$13.1$\pm$7.2) MeV and $X(2000)$ with mass M = (2002.1$\pm$27.5$\pm$15.0) MeV were observed in $J/\psi\rightarrow \phi \eta\eta^{'}$ at BESIII \cite{ssbar1}. They might be the candidates of $s\overline{s}$ states and needs more confirmation.
In a recent experimental study, the resonance $\phi(2170)$, listed in PDG \cite{PDG} was also confirmed by BESIII Collaboration \cite{ssbar4,ssbar5}. 
Also, the most precise resonance parameters for the state $h_{1}(1415)$ were also determined at BESIII \cite{ssbar6}.

Other than the experimental study of the kaon  and strangeonium family, it is very interesting to study these mesons theoretically as well. A vast amount of literature is available on the  spectroscopy of light mesons, using various theoretical approaches, including relativistic quark model \cite{Godfrey1985,L. Y. Xiao2019}, non-relativistic linear potential quark model  \cite{Q. Li2021}, QCD-motivated relativistic quark model  \cite{Ebert2009}, the Regge trajectory approach \cite{Badalian2019,Anisovich2000,Masjuan2012}.
In Refs. \cite{Vijande12005, Vijande2}, a detailed analysis of light-pseudoscalar and vector mesons is carried out and $q\overline{q}$ spectrum is studied in a  generalised constituent quark model.
In the recent article, the authors revisited the kaon spectrum in which the quark-antiquark interaction is based on the non-perturbative phenomena of dynamical chiral symmetry breaking and color confinement as well as the perturbative one-gluon exchange force \cite{U. T. Nieto2022}.  
The authors of Ref. \cite{P.-Lian2015}, gives an overview of the strangeonium states and their experimental status, and the experimental techniques are described to explore strangeonium with the BESIII detector. 
The mass spectrum of the kaon family is analysed by using the modified Godfrey-Isgur model with a color screening
effect in the Ref. \cite{C.Q. Pang2017}.


Since various theoretical approaches give different predictions for these mesons. Therefore, more comparison of computations with  the experimental measurements is required to identify the excited kaon and strangeonium mesons.The main aim of the present work is to systematically study the light strange mesons and the pure $s\overline{s}$ states and to discuss the possible spin-parity of the experimentally observed excited resonances, and predict the $J^{P}$ values of the resonances which are yet to be confirmed experimentally.
With the assumption of the existence of the quasilinear Regge trajectories, relations between Regge slope, intercept, and the meson masses have been extracted.  With the help of these derived relations, the ground state masses of strangeonium are computed. Further, the Regge parameters of kaon and strangeonium mesons are extracted to obtain the orbitally and radially excited state masses in both the $(J,M^{2})$ and $(n,M^{2})$ planes.


The paper is structured as follows. After the introduction, in Sec. II a detailed description of theoretical model is presented and the complete calculation of mass spectra of  kaon and strangeonium is given. Regge trajectories are drawn in both ($J,M^{2}$) and ($n,M^{2}$) planes. Sec. III is mostly devoted to the discussion of our obtained results and comparing our masses with those produced by other theoretical approaches. Finally, we concluded our work in Sec. IV.

\section{Theoretical Framework}

One of the most defining aspects of Regge phenomenology is the Regge trajectories which relates the mass and the spin of hadrons. Several theories have been put forth to explain the Regge trajectories. In 1978, Nambu \cite{Nambu1974,Nambu1979} proposed one of the simplest explanation for linear Regge trajectories by assuming the uniform  interactions between quark and antiquark pair which results in the formation of a strong flux tube. Light quarks rotating with the speed of light at radius $R$ at the end of the tube and the mass originating in this flux tube is estimated as \cite{Bali2001},

\begin{equation}
	\label{eq:1}
	M = 2\int_{0}^{R}\dfrac{\sigma}{\sqrt{1-\nu^{2}(r)}}dr = \pi\sigma R ,
\end{equation}
where $\sigma$ represent the string tension, i.e the mass density per unit length. $\nu(r)$ is the velocity as a function of the distance from the
center of the string. Similarly, the angular momentum of this flux tube is estimated as,
\begin{equation}
	\label{eq:2}
	J = 2\int_{0}^{R}\dfrac{\sigma r \nu(r)}{\sqrt{1-\nu^{2}(r)}}dr = \dfrac{\pi\sigma R^{2}}{2}+c^{'} ,
\end{equation}
From above relations one can also write,
\begin{equation}
	\label{eq:3}
	J = \dfrac{M^{2}}{2\pi\sigma} + c^{''} ,
\end{equation}
where $c^{'}$ and $c^{''}$ are the constants of integration. Hence we can say that, $J$ and $M^{2}$ are linearly related to each other.
Assuming the linear Regge trajectories for light mesons, one can write the most general form of  the linear Regge trajectories as \cite{Wei2008,JuhiPRD1,JuhiEPJA},
\begin{equation}
	\label{eq:4}
	J = \alpha(M) = a(0)+\alpha^{'} M^2 ,
\end{equation}
where $a(0)$ and $\alpha^{'}$ represent the intercept and slope of the trajectory, respectively.
For a meson multiplet, the Regge parameters for different quark constituents can be related by the following relations:
\begin{equation}
	\label{eq:5}
	a_{i \overline{i}}(0) + a_{j \overline{j}}(0) = 2a_{i \overline{j}}(0) ,	
\end{equation}
\begin{equation}
	\label{eq:6}
	\dfrac{1}{{\alpha^{'}}_{i \overline{i}}} + \dfrac{1}{{\alpha^{'}}_{j \overline{j}}} = \dfrac{2}{{\alpha^{'}}_{i \overline{j}}} ,	
\end{equation}\\
where $i$ and $j$ represent quark flavours. Equation (\ref{eq:5}), the additivity of intercepts was derived from the dual resonance model \cite{Kawarabayashi1969}, and was found to be satisfied in two dimensional QCD \cite{R. C. Brower1977},  and the quark bremsstrahlung model \cite{V.V. Dixit1979}. Eq. (\ref{eq:6}), the additivity of inverse slopes was derived in a model based on the topological expansion and the quark-antiquark string picture of hadrons \cite{A. B. Kaidalov1982}, it is also satisfied in the formal chiral limit and the heavy quark limit for both mesons and baryons \cite{L. Burakovsky1998}.
\\
Using equations (\ref{eq:4}) and (\ref{eq:5}) and after solving them we get, 
\begin{equation}
	\label{eq:7}
	\alpha^{'}_{i \overline{i}}M^{2}_{i \overline{i}}+\alpha^{'}_{j \overline{j}}M^{2}_{j \overline{j}}=2\alpha^{'}_{i \overline{j}}M^{2}_{i \overline{j}} .
\end{equation}
Combining Eqs. (\ref{eq:6}) and (\ref{eq:7}), two pairs of solutions are obtained in terms of slope ratios and mesons masses which are expressed as \cite{Wei2008},
	\begin{equation}
		\label{eq:8}
		\dfrac{\alpha^{'}_{j \overline{j}}}{\alpha^{'}_{i \overline{i}}}=\dfrac{1}{2M^{2}_{j \overline{j}}}\times[(4M^{2}_{i \overline{j}}-M^{2}_{i \overline{i}}-M^{2}_{j \overline{j}})
		\pm\sqrt{{{(4M^{2}_{i \overline{j}}-M^{2}_{i \overline{i}}-M^{2}_{j \overline{j}}})^2}-4M^{2}_{i \overline{i}}M^{2}_{j \overline{j}}}].
	\end{equation}
	and 
	\begin{equation}
		\label{eq:9}	
		\dfrac{\alpha^{'}_{i \overline{j}}}{\alpha^{'}_{i \overline{i}}}=\dfrac{1}{4M^{2}_{i \overline{j}}}\times[(4M^{2}_{i \overline{j}}+M^{2}_{i \overline{i}}-M^{2}_{j \overline{j}}) 
		\pm\sqrt{{{(4M^{2}_{i \overline{j}}-M^{2}_{i \overline{i}}-M^{2}_{j \overline{j}}})^2}-4M^{2}_{i \overline{i}}M^{2}_{j \overline{j}}}].
	\end{equation}
\\\\
Now using Eq. (\ref{eq:8}), we can extract the high-power mass equalities for mesons which is expressed as \cite{Wei2008},  

\begin{equation}
	\label{eq:10}
	\dfrac{\alpha^{'}_{j \overline{j}}}{\alpha^{'}_{i \overline{i}}}=	\dfrac{\alpha^{'}_{k \overline{k}}}{\alpha^{'}_{i \overline{i}}}\times	\dfrac{\alpha^{'}_{j \overline{j}}}{\alpha^{'}_{k \overline{k}}} .
\end{equation}
where $k$ can be any quark flavour. We have,
\\
	\begin{equation}
		\label{eq:11}
		\begin{split}
			\dfrac{[(4M^{2}_{i \overline{j}}-M^{2}_{i \overline{i}}-M^{2}_{j \overline{j}})+\sqrt{{{(4M^{2}_{i \overline{j}}-M^{2}_{i \overline{i}}-M^{2}_{j \overline{j}}})^2}-4M^{2}_{i \overline{i}}M^{2}_{j \overline{j}}}]}{2M^{2}_{j \overline{j}}} \\
			=\dfrac{[(4M^{2}_{i \overline{k}}-M^{2}_{i \overline{i}}-M^{2}_{k \overline{k}})+\sqrt{{{(4M^{2}_{i \overline{k}}-M^{2}_{i \overline{i}}-M^{2}_{k \overline{k}}})^2}-4M^{2}_{i \overline{i}}M^{2}_{k \overline{k}}}]/2M^{2}_{k \overline{k}}}{[(4M^{2}_{j \overline{k}}-M^{2}_{j \overline{j}}-M^{2}_{k \overline{k}})+\sqrt{{{(4M^{2}_{j \overline{k}}-M^{2}_{j \overline{j}}-M^{2}_{k \overline{k}}})^2}-4M^{2}_{j \overline{j}}M^{2}_{k \overline{k}}}]/2M^{2}_{k \overline{k}}}.
		\end{split}
	\end{equation}	
\\
In terms of different flavours of meson masses, we have derived a general relation expressed above, which can be used to evaluate the mass of any meson state if all other masses are known.

\subsection{Excited state masses of kaons and  strangeonium mesons in the ($J,M^{2}$) plane}	
In this section firstly we obtained the ground state ($1^{1}S_{0}$ and $1^{3}S_{1}$) masses of strangeonium meson. Since, experimentally it is very difficult to measure the masses of pure strangeonium state due to the usual mixing of pure $n\overline{n}$ and $s\overline{s}$ states. In the present work we do not consider the states with mixing of flavours.  Further the orbitally excited state masses of strangeonium and kaon are evaluated in the ($J,M^{2}$) plane using the above extracted relations. To determine the ground state masses we use relation (\ref{eq:11}). Since strangeonium meson composed of one strange and one anti-strange quark, hence we put $i=n(u$ or $d)$, $j=s$, and $k=c$ in the Eq. (\ref{eq:11}) and get the quadratic mass expression in terms of well established light and heavy mesons which is expressed as,

	\begin{equation}
		\label{eq:12}
		\begin{split}
			\dfrac{[(4M^{2}_{n \overline{s}}-M^{2}_{n \overline{n}}-M^{2}_{s \overline{s}})+\sqrt{{{(4M^{2}_{n \overline{s}}-M^{2}_{n \overline{n}}-M^{2}_{s \overline{s}}})^2}-4M^{2}_{n \overline{n}}M^{2}_{s \overline{s}}}]}{2M^{2}_{s \overline{s}}} \\
			=\dfrac{[(4M^{2}_{n \overline{c}}-M^{2}_{n \overline{n}}-M^{2}_{c \overline{c}})+\sqrt{{{(4M^{2}_{n \overline{c}}-M^{2}_{n \overline{n}}-M^{2}_{c \overline{c}}})^2}-4M^{2}_{n \overline{n}}M^{2}_{c \overline{c}}}]}{[(4M^{2}_{s \overline{c}}-M^{2}_{s \overline{s}}-M^{2}_{c \overline{c}})+\sqrt{{{(4M^{2}_{s \overline{c}}-M^{2}_{s \overline{s}}-M^{2}_{c \overline{c}}})^2}-4M^{2}_{s \overline{s}}M^{2}_{c \overline{c}}}]}.
		\end{split}
	\end{equation}	
By inserting the experimentally observed mass values of  $n\overline{n}$, $n\overline{s}$, $n\overline{c}$, $s\overline{c}$, and $c\overline{c}$ mesons for $J^{P}=0^{-}$ from PDG \cite{PDG} in the above relation, we get $M_{s\overline{s}}$ = 695.8 MeV. Similarly we can obtain the ground state mass for $J^{P}=1^{-}$ as 1005.6 MeV. 
\\\\
Now for the determination of orbitally excited state masses of kaons ($n\overline{s}$) and strangeonium ($s\overline{s}$), firstly we extract the values of Regge slopes ($\alpha^{'}$) using the relations (\ref{eq:8}) and (\ref{eq:9}) for these light flavoured mesons. For instance, according to the quark composition of kaons we insert $i=n$ and $j=s$ into Eq. (\ref{eq:9}), we get a expression in terms of slope ratios and light flavoured mesons, 

\begin{eqnarray}
	\label{eq:13} 
	\dfrac{\alpha^{'}_{n \overline{s}}}{\alpha^{'}_{n \overline{n}}}=\dfrac{1}{4M^{2}_{n \overline{s}}}\times[(4M^{2}_{n \overline{s}}+M^{2}_{n \overline{n}}-M^{2}_{s \overline{s}}) 
	+\sqrt{{{(4M^{2}_{n \overline{s}}-M^{2}_{n \overline{n}}-M^{2}_{s \overline{s}}})^2}-4M^{2}_{n \overline{n}}M^{2}_{s \overline{s}}}].
\end{eqnarray}
Since we have taken the experimental masses as inputs, 
we have also incorporated the experimental errors while calculating the Regge parameters and evaluating the excited state masses. 
Hence, we insert the values $M_{n\overline{n}}$ = 134.98$\pm$0.0005 MeV and $M_{n\overline{s}}$ = 497.61$\pm$0.013 from PDG \cite{PDG}, $M_{s\overline{s}}$ obtained above for $J^{P} = 0^{-}$ into the Eq. (\ref{eq:13}). Now, with the help of $\alpha^{'}_{n\overline{n}} = 2/(M_{\pi_{2}(1670)}-M_{\pi})$, we can get the value of  $\alpha^{'}_{n \overline{s}}$ as 0.7101$\pm$0.0001 GeV$^{-2}$ for $1^{1}S_{0}$ trajectory. Here we expressed our obtained Regge slopes as, $\alpha^{'}\pm\delta \alpha^{'}$, where $\delta \alpha^{'}$ is the experimental error in the slope. Similarly we can calculate the Regge slope, $\alpha^{'}_{n \overline{s}}$ = 0.8203$\pm$0.02 GeV$^{-2}$ for $1^{3}S_{1}$ trajectory. 
\\\\
In the same manner for strangeonium ($s\overline{s}$), which is composed of one strange quark and one anti-strange quark, we put $i=n$ and $j=s$ in Eq. (\ref{eq:8}) and get, 

\begin{eqnarray}
	\label{eq:14} 
	\dfrac{\alpha^{'}_{s \overline{s}}}{\alpha^{'}_{n \overline{n}}}=\dfrac{1}{2M^{2}_{s \overline{s}}}\times[(4M^{2}_{n \overline{s}}-M^{2}_{n \overline{n}}-M^{2}_{s \overline{s}}) 
	+\sqrt{{{(4M^{2}_{n \overline{s}}-M^{2}_{n \overline{n}}-M^{2}_{s \overline{s}}})^2}-4M^{2}_{n \overline{n}}M^{2}_{s \overline{s}}}].
\end{eqnarray}
Again inserting the values of $M_{n\overline{n}}$, $M_{n\overline{s}}$, and $M_{s\overline{s}}$ in the above relation and extracted the values of Regge slopes for $0^{-}$ and $1^{-}$ trajectories as 0.6750$\pm$0.001 GeV$^{-2}$ and 0.7721$\pm$0.032 GeV$^{-2}$ respectively in the similar manner.
From Eq. (\ref{eq:4}) an another relation can be obtained in terms of mesons masses and Regge slope which is expressed as,

\begin{equation}
	\label{eq:15}
	M_{J+1} = \sqrt{M_{J}^{2}+\dfrac{1}{\alpha^{'}}} .
\end{equation}
\\\\
Hence, with the help of Eq. (\ref{eq:15}) and extracted values of $\alpha^{'}_{n\overline{s}}$ and  $\alpha^{'}_{s\overline{s}}$, we can obtain the excited state masses of kaon and strangeonium mesons lying on $1^{1}S_{0}$  and $1^{3}S_{1}$ trajectories  for unnatural  ($1^{1}P_{1}$, $1^{1}D_{2}$, $1^{1}F_{3}$....) and natural ($1^{3}P_{2}$, $1^{3}D_{3}$, $1^{3}F_{4}$....) parity states.
Now, other than natural and unnatural parities, we have also calculate the other remaining states in the same manner. Firstly, using Eq. (\ref{eq:12}) we have obtained the masses for $1^{3}P_{0}$ and $1^{3}P_{1}$ strangeonium states as we have evaluated for the ground states earlier, by inserting the masses of other light and heavy mesons from PDG \cite{PDG} for $1^{3}P_{0}$ and $1^{3}P_{1}$ states. For some mesonic states, due to the unavailability of experimental masses we have taken the input masses from the Ref. \cite{Ebert2009}. Further, the Regge slopes for $1^{3}P_{0}$ and $1^{3}P_{1}$ trajectories have been extracted using the same procedure to calculate the higher excited states $1^{3}D_{1}$, $1^{3}F_{2}$, $1^{3}G_{4}$.... and $1^{3}D_{2}$, $1^{3}F_{3}$, $1^{3}G_{5}$... respectively lying on that trajectories.
We have include the experimental error in the calculated excited state masses. We expressed our obtained mass as, $M \pm \delta M$ where $M$ and $\delta M$ denotes the calculated mass and the experimental error in the mass respectively. The estimated results calculated in the $(J,M^{2})$ plane are represented in Tables \ref{tab:table1} and \ref{tab:table2} for kaons and strangeonium respectively, along with the comparison of experimental values where available and the predictions of other theoretical approaches. 
	
	\begingroup
\setlength{\tabcolsep}{8pt} 
\begin{table}
	\begin{center}
		\caption{
			Excited state masses of kaons in the $(J,M^{2})$ plane (in MeV). The numbers in the boldfaced are taken as inputs.
		}
		\label{tab:table1}
		\begin{tabular}{lllllllllllll}
			\hline
			\textit{$N^{2S+1}L_{J}$}&This work& Mesons&PDG \cite{PDG}& \cite{Ebert2009} & \cite{C.Q. Pang2017} & \cite{U. T. Nieto2022} &\cite{Godfrey1985} & \cite{Vijande12005}	 \\	
			\hline\noalign{\smallskip}

			$1^{1}S_{0}$ &\textbf{497.61$\pm$0.013}&$K^{\pm}$ &497.61$\pm$0.013&482&497.7&481 &462 &496 \\ 
			$1^{3}S_{1}$ &\textbf{891.67$\pm$0.26}&$K^{*}(892)$ &891.67$\pm$0.26 &897&896 &900 &903 &910 \\
			
			$1^{1}P_{1}$ &1286.81$\pm$0.77&$K_{1}(1270)$ &1253$\pm$7 &1294&1364 &1370 &1352 &1372\\ 
			$1^{3}P_{0}$ & 1362.00 $\pm$0.00 & & &1362 & 1257 & 1305 &1234 & 1394 \\
			$1^{3}P_{1}$ & 1403.00 $\pm$7.00 & & &1412 & 1377 &1455 &1366\\
			$1^{3}P_{2}$ &1419.20$\pm$10.47& $K^{*}_{2}(1430)$ &1425.6$\pm$1.5 &1424&1431 &1454 &1428 & 1450\\ 
			
			$1^{1}D_{2}$ &1750.46$\pm$0.80&$K_{2}(1770)$ &1773$\pm$8 &1709&1778 &1760 &1791 &1747 \\ 
			$1^{3}D_{1}$ & 1700.77$\pm$30.12 &$K^{*}(1680)$ &1718$\pm$18 & 1699 & 1766 &1787 & 1776 &1698\\
			$1^{3}D_{2}$ & 1756.84$\pm$61.44 &$K_{2}(1820)$ &1819$\pm$2 &1824 &1789 & 1854 &1804 &1741 \\
			$1^{3}D_{3}$ &1798.11$\pm$11.69& $K^{*}_{3}(1780)$ &1776$\pm$7&1789&1781 &1810 &1794 &1766 \\ 
			
			$1^{1}F_{3}$ &2114.79$\pm$0.81 & & &2009&2075 &2047 &2131\\
			$1^{3}F_{2}$ & 1982.46$\pm$36.54 &  & &1964 &2093 &2095 & 2151 &1968\\
			$1^{3}F_{3}$ & 2050.50$\pm$74.29 & & &2080 & 2084 & 2132 &2143\\
			$1^{3}F_{4}$ &2110.04$\pm$12.20&$K^{*}_{4}(2045)$ &2054$\pm$9 &2096&2058 &2080 &2108  \\ 
			
			$1^{1}G_{4}$ &2424.99$\pm$0.82 & &&2255&2309 &2270 &2422\\ 
			$1^{3}G_{3}$ &2228.84$\pm$39.81 & & &2207 &2336 &2316 & 2458\\
			$1^{3}G_{4}$ &2307.08$\pm$80.81 & & &2285 & 2317 & 2337 &2433\\
			$1^{3}G_{5}$ &2381.46$\pm$12.48& $K^{*}_{5}(2380)$ &2382$\pm$24 &2356 &2286 &2291 &2388\\ 
			
			$1^{1}H_{5}$ &2699.79$\pm$0.82 & & & & & 2442 \\ 
			$1^{3}H_{4}$ &2450.57$\pm$41.81& & & & &2477\\
			$1^{3}H_{5}$ &2537.85$\pm$84.80& & & & &2489\\
			$1^{3}H_{6}$ &2624.96$\pm$12.66\\ 
			
			\hline
			
		\end{tabular}
	\end{center}
\end{table}
\endgroup

\begingroup
\setlength{\tabcolsep}{8pt} 
\begin{table}
	\begin{center}
		\caption{
			Excited state	masses of strangeonuim in the $(J,M^{2})$ plane (in MeV). 
		}
		\label{tab:table2}
		\begin{tabular}{lllllllllllll}
			\hline
			\textit{$N^{2S+1}L_{J}$}&This work& Mesons&PDG \cite{PDG}&  \cite{Ebert2009} & \cite{Q. Li2021}& \cite{L. Y. Xiao2019} &\cite{S. Ishida1987} & \cite{Vijande12005}	 \\	
			\hline\noalign{\smallskip}

			$1^{1}S_{0}$ & 695.80$\pm$0.00 & & &743 &797 &675 &690 &956 \\ 
			$1^{3}S_{1}$ & 1005.63$\pm$0.00&$\phi$ &1019.46$\pm$0.016 & 1038 &1017 &1009 &1020 &1020\\
			
			$1^{1}P_{1}$ & 1402.01$\pm$0.78&$h_{1}(1415)$ &1416$\pm$8 &1485 &1462 &1473 &1460 &1511\\ 
			$1^{3}P_{0}$ & 1392.00$\pm$0.00 &$f_{0}(1370)$ & 1350$\pm$9$^{+12}_{-2}$ \cite{ssbar2} & 1420 &1373 &1355&1180 & 1340\\
			$1^{3}P_{1}$ & 1514.90$\pm$0.00 & & &  1464&1492&1480&1430&1508\\
			$1^{3}P_{2}$ & 1518.70$\pm$17.67& $f_{2}^{'}(1525)$ &1517.4$\pm$2.5 & 1529 &1513 &1539&1480 &1556\\
			
			$1^{1}D_{2}$ & 1856.64$\pm$0.83&$\eta_{2}(1870)$ &1842$\pm$8 &1909 &1825 &1893 &1830 & 1853\\ 
			$1^{3}D_{1}$ & 1731.50$\pm$30.24 & & &1845&1809&1883&1750&\\
			$1^{3}D_{2}$ & 1848.07$\pm$95.98 & & &1908&1840&1904&1810\\
			$1^{3}D_{3}$ & 1897.79$\pm$19.99 &$\phi_{3}(1850)$&1854$\pm$7 &1950 &1822 &1897 &1830 &1875\\ 
			
			$1^{1}F_{3}$ & 2220.04$\pm$0.85& & &2209 &2111 &2223 &2130\\ 
			$1^{3}F_{2}$ & 2014.59$\pm$36.75& & &2143&2146&2243 &2090\\
			$1^{3}F_{3}$ & 2129.74$\pm$117.78& & &2215&2128&2234 &2120\\	
			$1^{3}F_{4}$ & 2212.87$\pm$21.01& $f_{4}(2210)$ & 2209$_{-15}^{+17}$ \cite{LASS1988} & 2286 &2078 &2202 &2130\\
			
			$1^{1}G_{4}$ & 2531.81$\pm$0.86& & &2469 & &2507 \\
			$1^{3}G_{3}$ &2262.52$\pm$40.08& & &2403\\
			$1^{3}G_{4}$ &2378.29$\pm$129.18& & &2481\\
			$1^{3}G_{5}$ & 2488.37$\pm$21.57 && & 2559&\\
			
			$1^{1}H_{5}$ & 2809.19$\pm$0.87& & &2706  \\  
			$1^{3}H_{4}$ & 2485.85$\pm$42.12& & &2634 \\
			$1^{3}H_{5}$ & 2603.21$\pm$136.28& & &2720\\
			$1^{3}H_{6}$ & 2736.26$\pm$21.93& & & 2809&\\

			\hline
			
		\end{tabular}
	\end{center}
\end{table}
\endgroup

	\subsection{Excited state masses of kaons and strangeonium mesons in ($n,M^{2}$) plane}


%
%
After obtaining the orbitally excited state masses of kaon and strangeonium mesons, in this section the Regge parameters have been extracted and radial excitations are obtained in the ($n,M^{2}$) plane. 
The general equation for linear Regge trajectories in the ($n,M^{2}$) plane can be expressed as,
\begin{equation}
	\label{eq:16}
	n = \beta_{0} + \beta M^{2},
\end{equation}
where $n$ represents the radial principal quantum number; 1,2,3.... $\beta_{0}$, and $\beta$ are the intercept and slope of the trajectories. The meson multiplets lying on the single Regge trajectory have the same Regge slope ($\beta$) and Regge intercept ($\beta_{0}$). Using relation (\ref{eq:16}), the values of $\beta$ and $\beta_{0}$ for these light mesons can be extracted for $S$, $P$, $D$... states to evaluate the excited state masses lying on Regge trajectories. For instance, for strange mesons (kaons), with the help of slope equation we can have  $\beta_{(S)} = 1/(M^{2}_{K(2S)}-M^{2}_{K(1S)})$.
Here the mass values of first radial excitations ($n$=2)  are taken as inputs from PDG \cite{PDG}, the experimentally observed masses wherever available. Whereas, due to the unavailability of experimental masses in some states we have taken the theoretical predictions of \cite{Ebert2009} for the calculation. Here also, the error analysis is incorporated for the calculation of Regge parameters. 
Hence, by putting the values of $M_{K(1S)}$ and $M_{K(2S)}$ for $J^{P}=0^{-}$, we can get $\beta_{(S)}$ = 0.51285$\pm$0.003 GeV$^{-2}$ for $S$-states with spin $s$=0. Now from relation (\ref{eq:16}) we can write, 

\begin{equation}
	\label{eq:17}
	\begin{split}
		1 = \beta_{0(S)} + \beta_{(S)} M^{2}_{K(1S)},\\
		2 = \beta_{0(S)} + \beta_{(S)} M^{2}_{K(2S)},
	\end{split}
\end{equation}
\\\\
Simplifying the above equations, we get $\beta_{0(S)}$ = 0.87301$\pm$0.0007. With the help of $\beta_{(S)}$ and $\beta_{0(S)}$, we can evaluate the masses of the excited $K$-meson states for $n$ = 3, 4, 5... Similarly, we can express these relations for $P$ and $D$-wave as,
\begin{equation}
	\label{eq:18}
	\begin{split}
		1 = \beta_{0(P)} + \beta_{(P)} M^{2}_{K(1P)},\\
		2 = \beta_{0(P)} + \beta_{(P)} M^{2}_{K(2P)},\\
		1 = \beta_{0(D)} + \beta_{(D)} M^{2}_{K(1D)},\\
		2 = \beta_{0(D)} + \beta_{(D)} M^{2}_{K(2D)},
	\end{split}
\end{equation}
\\\\
With the help of above obtained relations we have extracted the values of 
Regge slopes and intercepts for kaon and strangeonium mesons for each Regge trajectory with spin $s$= 0 and 1, and obtain the radially excited state masses lying on that trajectory. The predicted mass spectra for kaon and strangeonium are shown in Tables \ref{tab:table3} and \ref{tab:table4} respectively along with the predictions of other theoretical approaches. 

	\begingroup
\setlength{\tabcolsep}{14pt} 
\begin{table}
	\small
	\begin{center}
		\caption{
			Excited state masses of kaons in the $(n,M^{2})$ plane (in MeV). The numbers in the boldfaced are taken as inputs.
		}
		\label{tab:table3}
		
		\begin{tabular}{lllllllllllll}
			\hline
			\textit{$N^{2S+1}L_{J}$}&This work& \cite{U. T. Nieto2022} & \cite{C.Q. Pang2017} & \cite{Godfrey1985}	& \cite{Vijande12005}   \\
			\hline\noalign{\smallskip}
			
			$1^{1}S_{0}$ & 497.61$\pm$0.013& 481 &497.7  &462 &496\\ 
			$2^{1}S_{0}$ & \textbf{1482.40$\pm$3.60} \cite{PDG} & 1512 &1457 &1454 &1472\\ 
			$3^{1}S_{0}$ & 2036.52$\pm$5.57 & 2018 &1924 &2065 &1899\\ 
			$4^{1}S_{0}$ & 2469.27$\pm$6.74 & 2318 &2248\\ 
			$5^{1}S_{0}$ & 2836.76$\pm$7.75 & 2488\\ 
				$6^{1}S_{0}$ & 3161.82$\pm$8.63 & 2467\\ 
			\noalign{\smallskip}
			$1^{3}S_{1}$ & 891.67$\pm$0.26 & 900 &896 &903 &910\\ 
			$2^{3}S_{1}$ & \textbf{1675.00$\pm$0.00} \cite{Ebert2009}  & 1676 &1548 &1579 &1620\\ 
			$3^{3}S_{1}$ & 2194.57$\pm$0.24&2112 &1983 &1950\\ 
			$4^{3}S_{1}$ & 2612.79$\pm$0.27&2372 &2287\\ 
			$5^{3}S_{1}$ & 2972.75$\pm$0.31&2516\\ 
					$6^{3}S_{1}$ & 3293.59$\pm$0.34 &2576\\ 
			\hline\noalign{\smallskip}
			$1^{1}P_{1}$ & 1286.81$\pm$0.77 &1370 &1364 &1352 &1372\\
			$2^{1}P_{1}$ & \textbf{1757.00$\pm$0.00}\cite{Ebert2009}     &1925 &1840 &1897 &1841\\
			$3^{1}P_{1}$ & 2125.61$\pm$1.66 &2260 &2177 &2164\\
			$4^{1}P_{1}$ & 2439.14$\pm$1.84  &2458 &2422\\
				$5^{1}P_{1}$ & 2716.72$\pm$2.01  &2556\\  
			\noalign{\smallskip}
			$1^{3}P_{0}$ & 1362.00$\pm$0.00 &1305 & 1257 & 1234 & 1213\\
			$2^{3}P_{0}$ & \textbf{1791.00$\pm$0.00}\cite{Ebert2009} &1894 & 1829 & 1890 & 1768\\
			$3^{3}P_{0}$ & 2135.49$\pm$0.00 & 2242 &2176 & 2160\\
			$4^{3}P_{0}$ & 2431.66$\pm$0.00 & 2447& 2424\\
				$5^{3}P_{0}$ & 2695.48$\pm$0.00 & 2552\\
			\noalign{\smallskip}
			$1^{3}P_{1}$ & 1403.00$\pm$7.00 & 1455 & 1377 &1366 & 1394\\
			$2^{3}P_{1}$ & \textbf{1893.00$\pm$0.00}\cite{Ebert2009} &1971 & 1861 & 1928 & 1850\\
			$3^{3}P_{1}$ & 2280.11$\pm$15.44 &2286 & 2192 & 2200\\
			$4^{3}P_{1}$ & 2610.39$\pm$16.95 & 2471 & 2434\\
				$5^{3}P_{1}$ & 2903.34$\pm$18.40 &2561\\
			\noalign{\smallskip}
			$1^{3}P_{2}$ & 1419.20$\pm$10.47 &1454 &1431 &1428 &1450\\
			$2^{3}P_{2}$ & \textbf{1994.00$\pm$60.00}\cite{PDG}   &1975 &1870 &1938\\
			$3^{3}P_{2}$ & 2436.79$\pm$158.22&2290 &2198 &2206\\
			$4^{3}P_{2}$ & 2810.66$\pm$178.30&2474 &2438\\
					$5^{3}P_{2}$ & 3140.34$\pm$197.00&2563\\ 
			\hline\noalign{\smallskip}
			
			$1^{1}D_{2}$ & 1750.46$\pm$0.80 &1760 &1778 &1791 &1747\\ 
			$2^{1}D_{2}$ &\textbf{2066.00$\pm$0.00} \cite{Ebert2009}   &2160 &2121 &2238\\ 
			$3^{1}D_{2}$ & 2339.35$\pm$3.17&2402 &2380\\ 
			$4^{1}D_{2}$ & 2583.95$\pm$3.35&2535 &2570\\ 
				$5^{1}D_{2}$ & 2807.31$\pm$3.53&2588\\ 
			\noalign{\smallskip}
			$1^{3}D_{1}$ & 1700.77$\pm$30.12 & 1787 & 1766 &1776 & 1698\\
			$2^{3}D_{1}$ & \textbf{2063.00$\pm$0.00}\cite{Ebert2009} & 2173 & 2127 & 2251\\
			$3^{3}D_{1}$ & 2370.51$\pm$102.48 & 2408 & 2385 &\\
			$4^{3}D_{1}$ & 2642.47$\pm$109.20 & 2537 & 2573\\
				$5^{3}D_{1}$ & 2888.95$\pm$123.27 & 2585\\ 
			\noalign{\smallskip}
			$1^{3}D_{2}$ & 1756.84$\pm$61.44 & 1854 & 1789 &1804 &1741\\
			$2^{3}D_{2}$ & \textbf{2163.00$\pm$0.00}\cite{Ebert2009} & 2214 & 2131 &  2254\\
			$3^{3}D_{2}$ & 2504.13$\pm$194.07 &2432 & 2388 &\\
			$4^{3}D_{2}$ & 2804.06$\pm$207.83 &2547 & 2575\\
				$5^{3}D_{2}$ & 3074.87$\pm$222.09 & 2588\\
			\noalign{\smallskip}
			$1^{3}D_{3}$ & 1798.11$\pm$11.69&1810 &1781 &1794 &1766\\
			$2^{3}D_{3}$ & \textbf{2182.00$\pm$0.00}\cite{Ebert2009}   &2191 &2131 &2237\\ 
			$3^{3}D_{3}$ & 2507.80$\pm$39.67 &2421 &2382\\
			$4^{3}D_{3}$ & 2795.88$\pm$42.28&2543 &2571\\
				$5^{3}D_{3}$ & 3056.94$\pm$45.01&2587\\
			
			\hline

		\end{tabular}
	\end{center}
\end{table}
\endgroup

\begingroup
\setlength{\tabcolsep}{10pt} 

\begin{table}
	\small
	\begin{center}
		\caption{
			Masses of excited states of the strangeonium  in the $(n,M^{2})$ plane (in MeV). The numbers in the boldfaced are taken as inputs.
		}
		\label{tab:table4}
		\begin{tabular}{lllllllllllll}
			\hline
			\textit{$N^{2S+1}L_{J}$}&This work&PDG\cite{PDG}&\cite{L. Y. Xiao2019} & \cite{Q. Li2021}  & \cite{S. Ishida1987}	& \cite{Godfrey1985} & \cite{Vijande12005} & \\	
			\hline\noalign{\smallskip}
			
			$1^{1}S_{0}$ & 695.80$\pm$0.00 & &657 &797 &690 &960 &956\\ 
			$2^{1}S_{0}$ & \textbf{1475.00$\pm$4.00} \cite{PDG} &1475$\pm$4 &1578 &1619 &1440 &1630 &1795\\ 
			$3^{1}S_{0}$ & 1966.51$\pm$6.70& &2125 &2144 &1970\\ 
			$4^{1}S_{0}$ & 2357.68$\pm$8.25 &&2568 &2580 &2260\\ 
			$5^{1}S_{0}$ & 2692.61$\pm$9.40 &&2949\\ 
				$6^{1}S_{0}$ & 2990.26$\pm$10.43 & & 3328\\ 
			\noalign{\smallskip}
			$1^{3}S_{1}$ & 1005.63$\pm$0.00&1019.46$\pm$0.016 &1009 &1017 &1020 &1020 &1020 \\ 
			$2^{3}S_{1}$ & \textbf{1680.00$\pm$20.00}\cite{PDG}& &1688 &1699 &1740 &1690 &1726\\ 
			$3^{3}S_{1}$ & 2152.56$\pm$39.98 &2162$\pm$7&2204&2198 &2250 & & &\\ 
			$4^{3}S_{1}$ & 2538.62$\pm$46.53&&2627 &2623 &2540 & & &\\ 
			$5^{3}S_{1}$ & 2873.20$\pm$52.42&&2996\\ 
				$6^{3}S_{1}$ & 3172.83$\pm$57.74 & & 3327\\ 
			
			\hline\noalign{\smallskip}
			$1^{1}P_{1}$ &1402.01$\pm$0.78&1416$\pm$8 &1473 &1462 &1460 &1470 &1511\\
			$2^{1}P_{1}$ &\textbf{2024.00$\pm$0.00}\cite{Ebert2009}&&2008 &1991 &2040 &2010 &1973\\
			$3^{1}P_{1}$ & 2495.51$\pm$1.46& &2449 &2435 &2490\\
			$4^{1}P_{1}$ & 2891.11$\pm$1.62 &&2832\\
				$5^{1}P_{1}$ & 3238.75$\pm$1.78 &&3174\\  
			\noalign{\smallskip}
			$1^{3}P_{0}$ & 1392.00$\pm$0.00 & &1355 &1373 &1180 &1360 &1340\\
			$2^{3}P_{0}$ & \textbf{1909.00$\pm$0.00}\cite{Ebert2009} & &1986 &1971 &1800 &1990 &1894\\
			$3^{3}P_{0}$ & 2411.69$\pm$0.00 & 2411$\pm$10$\pm7$ \cite{LASS1988} &2444 &2434 &2280\\
			$4^{3}P_{0}$ & 2784.88$\pm$0.00 & &2834\\
				$5^{3}P_{0}$ & 3113.66$\pm$0.00 & &3179\\
			\noalign{\smallskip}
			$1^{3}P_{1}$ & 1514.90$\pm$0.00 & &1480 &1492 &1430 &1480 &1508\\
			$2^{3}P_{1}$ & \textbf{2016.00$\pm$0.00}\cite{Ebert2009} & &2027 &2027 &2020 &2030\\
			$3^{3}P_{1}$ & 2415.28$\pm$0.00& &2468&2470 &2480\\
			$4^{3}P_{1}$ & 2757.34$\pm$0.00 & &2850 &\\
				$5^{3}P_{1}$ & 3061.41$\pm$0.00 & &3191\\
			\noalign{\smallskip}
			$1^{3}P_{2}$ & 1518.70$\pm$17.67& 1517.4$\pm$2.5&1539 &1513 &1480 &1530 &1556\\
			$2^{3}P_{2}$ & \textbf{2011.00$\pm$60.00}\cite{PDG}& &2046 &2030 &2080 &2040 &1999\\
			$3^{3}P_{2}$ & 2404.54$\pm$184.30 &&2480 &2466 &2540\\
			$4^{3}P_{2}$ & 2742.16$\pm$204.13 &&2859\\
				$5^{3}P_{2}$ & 3042.55$\pm$223.12 &&3198\\ 
			\hline\noalign{\smallskip}
			$1^{1}D_{2}$ & 1856.64$\pm$0.83& 1842$\pm$8&1893 &1825 &1830 &1890 &1853\\ 
			$2^{1}D_{2}$ & \textbf{2321.00$\pm$0.00}\cite{Ebert2009}& &2336 &2282 &2340\\ 
			$3^{1}D_{2}$ & 2706.84$\pm$2.42& &2723 &2685\\ 
			$4^{1}D_{2}$ & 3044.16$\pm$2.60& &3070\\ 
				$5^{1}D_{2}$ & 3347.66$\pm$2.79& &3387\\ 
			\noalign{\smallskip}
			$1^{3}D_{1}$ & 1731.50$\pm$30.24 & &1883 &1809 &1750 &1880\\
			$2^{3}D_{1}$ & \textbf{2258.00$\pm$0.00}\cite{Ebert2009} & &2342 &2272 &2260 &\\
			$3^{3}D_{1}$ & 2683.10$\pm$75.03& &2732 &2681\\
			$4^{3}D_{1}$ & 3049.51$\pm$81.69 & &3079\\
				$5^{3}D_{1}$ & 3376.39$\pm$88.39& &3395\\
			\noalign{\smallskip}
			$1^{3}D_{2}$ & 1848.07$\pm$95.98 & &1904 &1840 &1810 &1910\\
			$2^{3}D_{2}$ & \textbf{2323.00$\pm$0.00}\cite{Ebert2009} & &2348 &2297 &2330\\
			$3^{3}D_{2}$ & 2716.12$\pm$275.84 & &2734 &2701\\
			$4^{3}D_{2}$ & 3059.13$\pm$297.30 & &3080\\
				$5^{3}D_{2}$ & 3367.38$\pm$319.27 & &3396\\
			\noalign{\smallskip}
			$1^{3}D_{3}$ & 1897.79$\pm$19.99 &1854$\pm$7&1897 &1822 &1830 &1900 &1875\\
			$2^{3}D_{3}$ & \textbf{2338.00$\pm$0.00}\cite{Ebert2009}& &2337 &2285 &2360\\ 
			$3^{3}D_{3}$ & 2707.56$\pm$62.94& &2725 &2691\\
			$4^{3}D_{3}$ & 3032.41$\pm$67.42& &3073\\
				$5^{3}D_{3}$ & 3325.69$\pm$72.06& &3390\\
			\hline
			
		\end{tabular}
	\end{center}
\end{table}
\endgroup

\subsection{Regge trajectories in the ($J,M^{2}$) and $(n,M^{2})$ planes.}

\begin{figure*}
	\centering
	\includegraphics[scale=0.3]{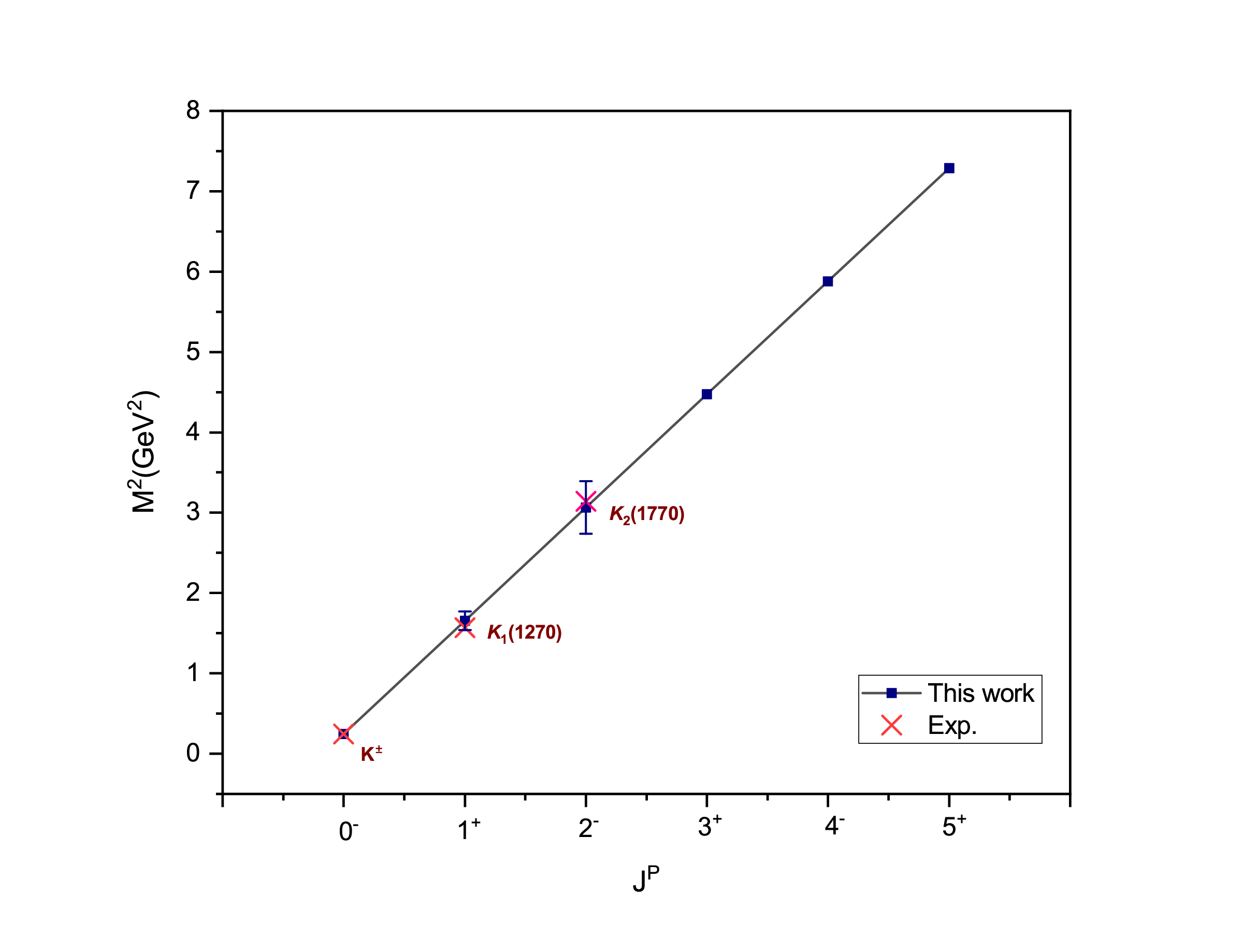}
	\caption{\label{fig:1}{Regge trajectory in the ($J,M^{2}$) plane for kaon with unnatural parity states. Error bars correspond to take $\Delta M^{2} = \pm \Gamma M$.}}
\end{figure*}

\begin{figure*}
	\centering
	\includegraphics[scale=0.3]{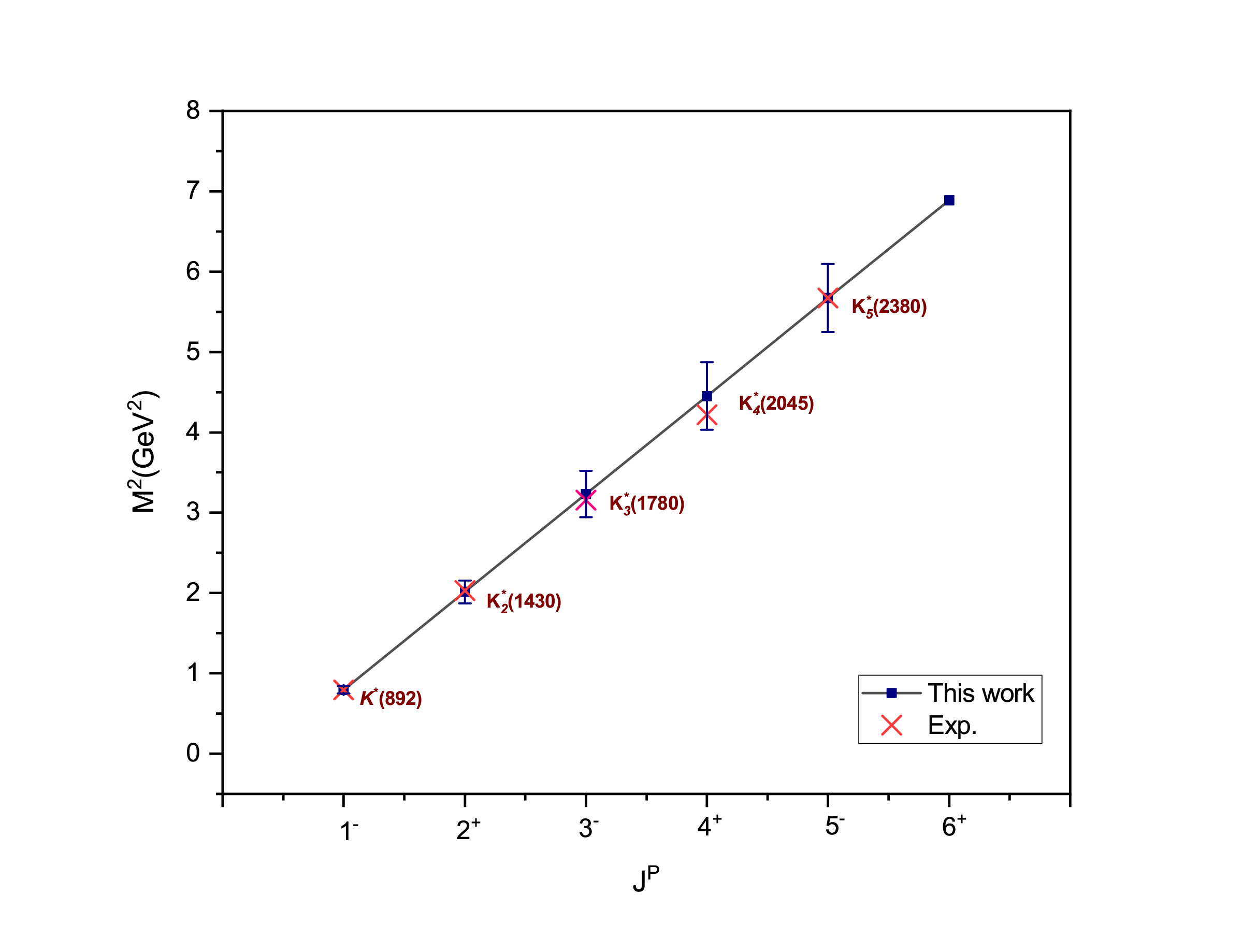}
	\caption{\label{fig:2}{Regge trajectory in the ($J,M^{2}$) plane for kaon with natural parity states. Error bars correspond to take $\Delta M^{2} = \pm \Gamma M$.}}
\end{figure*}

\begin{figure*}
	\centering
	\includegraphics[scale=0.3]{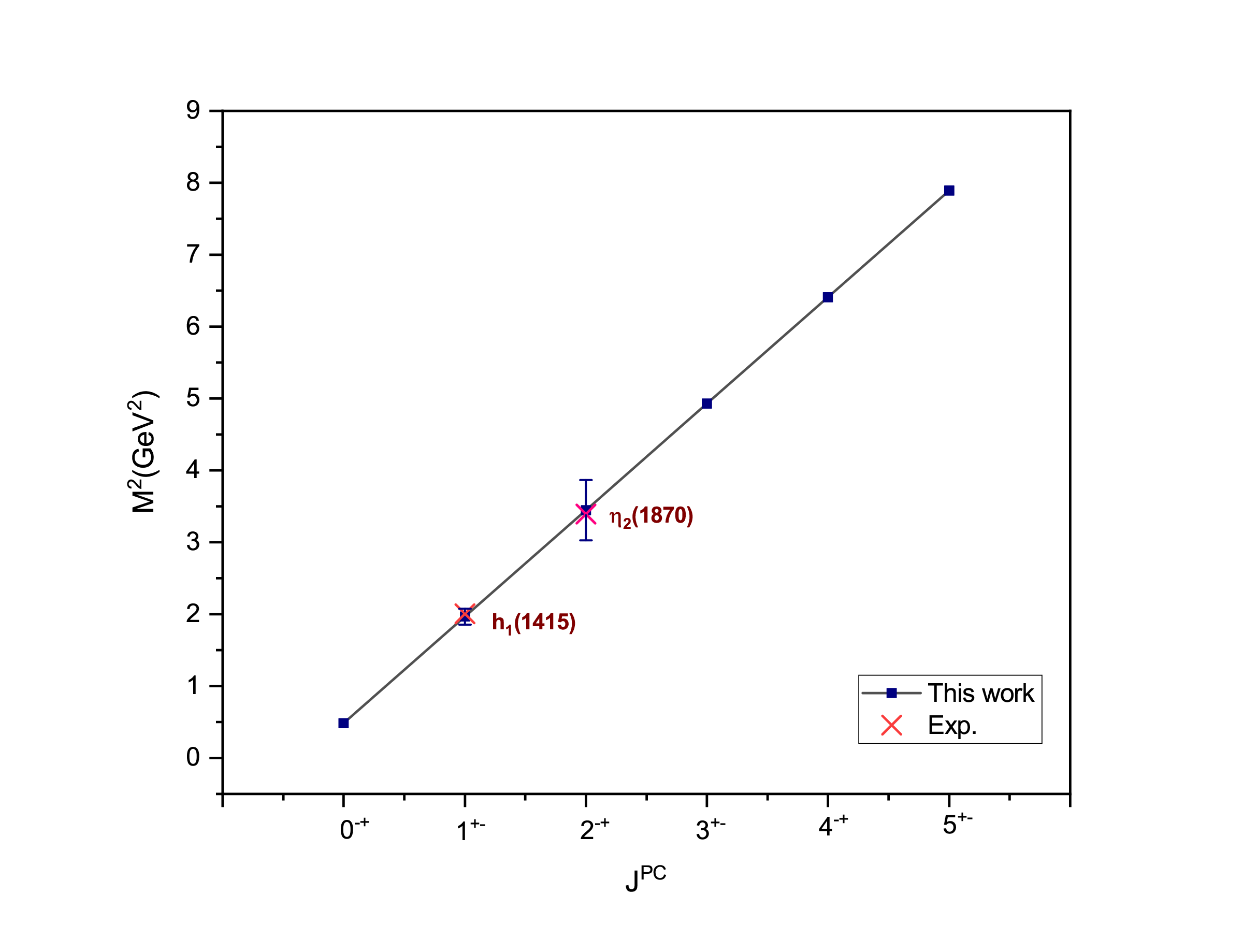}
	\caption{\label{fig:3}{Regge trajectory in the ($J,M^{2}$) plane for strangeonium with unnatural parity states. Error bars correspond to take $\Delta M^{2} = \pm \Gamma M$.}}
\end{figure*}

\begin{figure*}
	\centering
	\includegraphics[scale=0.3]{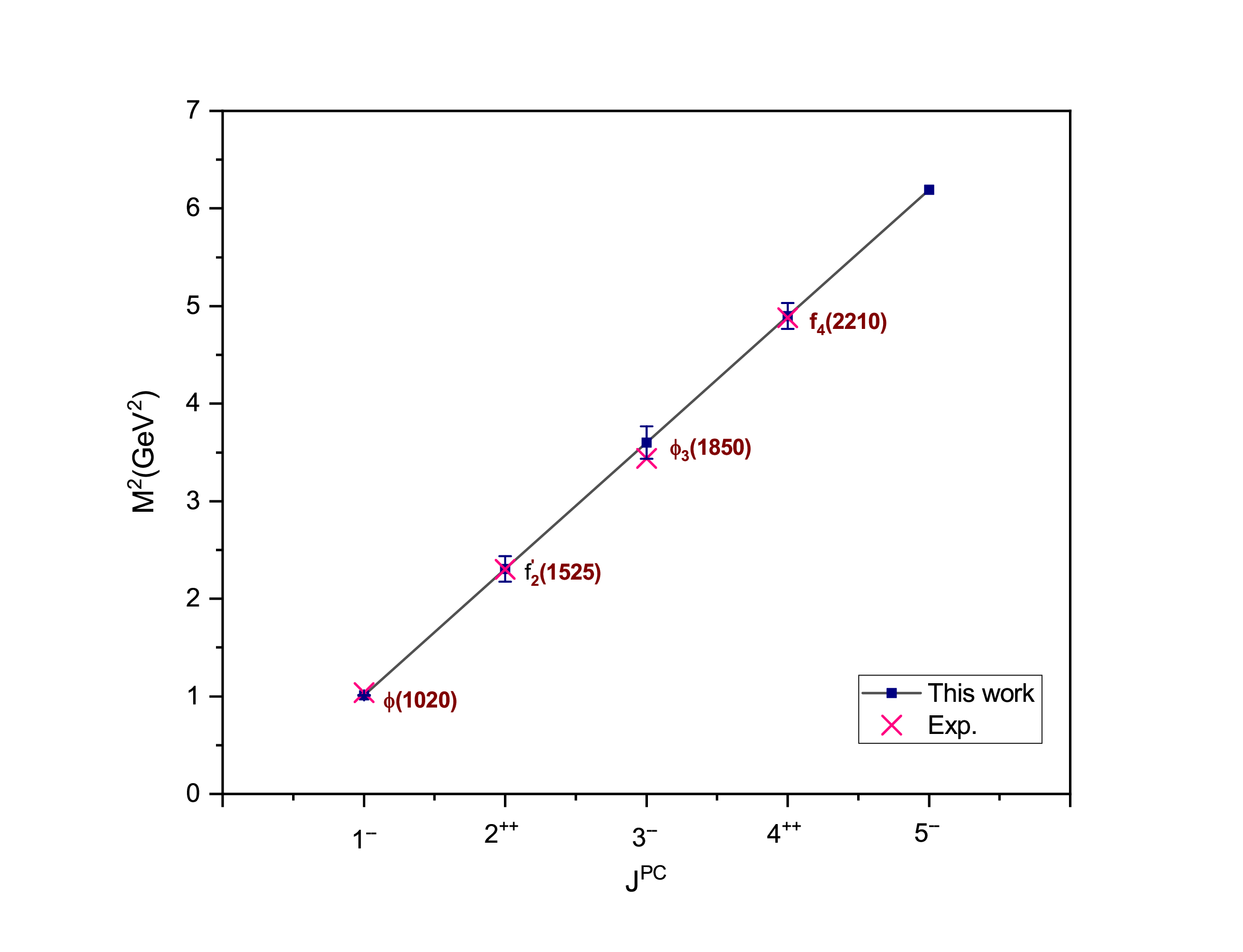}
	\caption{\label{fig:4}{Regge trajectory in the ($J,M^{2}$) plane for strangeonium with natural parity states. Error bars correspond to take $\Delta M^{2} = \pm \Gamma M$.}}
\end{figure*}

In this section, based on the calculated results of kaons and strangeonium mesons, the plots of total angular momentum, $J$, and principal quantum number, $n$, against the square of resonance mass $M^{2}$ are constructed. Additionally, in our all $M^{2}$ plots we have also incorporated the error which  is given by $\Delta M^{2}$ = $\pm \Gamma M$, where $\Gamma$ represents the width of the resonance. Hence, we have taken the width of each state as an estimate of the error of the resonance mass and this mass uncertainty is referred as  half width rule \cite{Masjuan2012,E. Ruiz2011,E. Ruiz2010}. The squared mass of each meson is then represented as $M^{2} \pm \Gamma M$. The numerical values of the resonance widths are the experimental data which are taken from the latest PDG \cite{PDG}. Tables \ref{tab:tablea} and \ref{tab:tableb} represents the estimated mass uncertainty $\Delta M^{2} = \pm \Gamma M$ of the states belongs to kaons and strangeonium mesons respectively.
\\\\
The Regge trajectories in the ($J,M^{2}$) plane are drawn for kaon and strangeonium with unnatural and natural parity states as shown in Figs. \ref{fig:1} - \ref{fig:4}. The solid straight lines represents our calculated results and the cross symbol represents the experimental masses. Also, the error bars represents the uncertainty. Similarly, the trajectories in the ($n,M^{2}$) plane are depicted in Figs. \ref{eq:5}-\ref{eq:8}.  
\\\\
\begin{table}[h!]
	\begin{center}
		\caption{
			The uncertainty in the squared masses of kaons which is given by 	$\Delta M^{2}$ = $\pm \Gamma M$. The numerical values of widths ($\Gamma$) are the experimental values taken from PDG \cite{PDG}.
		}
		\label{tab:tablea}
		\begin{tabular}{cccccccccccccccccccccccccc}
			\hline 
			\textit{$N^{2S+1}L_{J}$} & Resonances & Exp. Mass \cite{PDG} &  Mass M (Ours) &Width ($\Gamma$) \cite{PDG} &	$\Delta M^{2}$ = $\pm \Gamma M$ \\
			& & (MeV) & (MeV) & (MeV) & (GeV$^{2}$) \\
			\hline\noalign{\smallskip}
			$1^{3}S_{1}$ & K(892) & 891.67 & 891.67 & 51.4 & 0.0458 \\
			$2^{1}S_{0}$ &$K(1460)$	& 1482.40 & 1482.40 &335.60 &0.4975 \\
			$1^{1}P_{1}$ & $K_{1}(1270)$ &1253.00 &1286.81 & 90 &0.1158 \\
			$1^{3}P_{2}$ & $K_{2}^{*}(1430)$ &1425.60 & 1419.20 & 100 &0.1419 \\
			$2^{3}P_{2}$ & $K_{2}^{*}(1980)$ & 1994.00 & 1994.00 &348 & 0.6939\\
			$1^{1}D_{2}$ & $K_{2}(1770)$ & 1773.00 & 1750.46 & 186 & 0.3256 \\
			$1^{3}D_{3}$ & $K_{3}^{*}(1780)$ & 1776.00 & 1798.11 & 161 & 0.2895 \\	
			$1^{3}F_{4}$ & $K_{4}^{*}(2045)$ & 2054.00 & 2110.04 & 199 & 0.4199 \\
			$1^{3}G_{5}$ & $K_{5}^{*}(2380)$ & 2382.00 & 2381.46 & 178 & 0.4239 \\

			\hline
			
		\end{tabular}
	\end{center}
\end{table}
\begin{table}[h!]
	\begin{center}
		\caption{
			The uncertainty in the squared masses of strangeonium which is given by 	$\Delta M^{2}$ = $\pm \Gamma M$. The numerical values of widths ($\Gamma$) are the experimental values taken from PDG \cite{PDG}.
		}
		
		\label{tab:tableb}
		
		\begin{tabular}{cccccccccccccccccccccccccc}
			\hline 
			\textit{$N^{2S+1}L_{J}$} & Resonances & Exp. Mass &  Mass M (Ours) &Width ($\Gamma$) \cite{PDG} &	$\Delta M^{2}$ = $\pm \Gamma M$ \\
			& & (MeV) & (MeV) & (MeV) & (GeV$^{2}$) \\
			\hline\noalign{\smallskip}
			$1^{3}S_{1}$ & $\phi (1020)$ & 1019.46 & 1005.63 & 4.25 &0.0043\\
			$2^{1}S_{0}$ & $\eta(1475)$ &1475.00 &1475.00 &90 &0.1328\\
			$2^{3}S_{1}$ & $\phi (1680)$ &1680.00 & 1680.00 & 150 &0.2520 \\
			$3^{3}S_{1}$ & $\phi (2170)$ & 2162.00 & 2152.56 & 103 & 0.2217 \\ 
			$1^{1}P_{1}$ & $h_{1}(1415)$ &1416.00 & 1402.01 & 78 &0.1094\\
			$1^{3}P_{2}$ & $f_{2}^{'}(1525)$ & 1517.40 & 1518.70 & 86 &0.1306\\
			
			$1^{1}D_{2}$ & $\eta_{2}(1870)$ & 1842.00 &1856.64 & 225 &0.4177\\
			$1^{3}D_{3}$ & $\phi_{3}(1850)$ &1854.00 &1897.79 & 87 &0.1651\\	
			$1^{3}F_{4}$ & $f_{4}(2210)$ & 2209.00 & 2212.87 & 60 &0.1328\\

			\hline
			
		\end{tabular}
	\end{center}
\end{table}


\begin{figure*}
	\centering
	\includegraphics[scale=0.3]{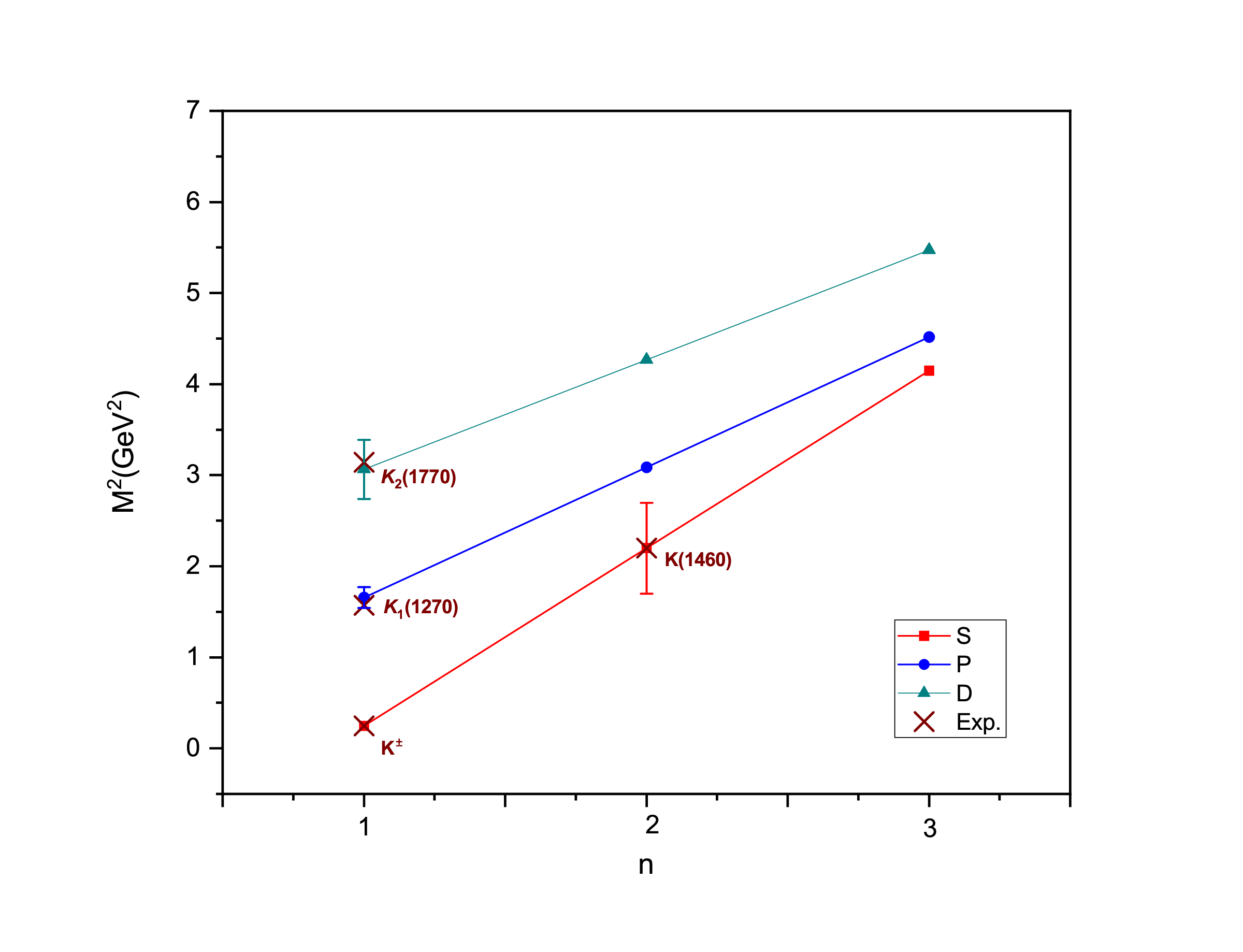}
	\caption{\label{fig:5}{Regge trajectories in the ($n,M^{2}$) plane for kaon with unnatural parity states. Error bars correspond to take $\Delta M^{2} = \pm \Gamma M$.}}
\end{figure*}
\begin{figure*}
	\centering
	\includegraphics[scale=0.3]{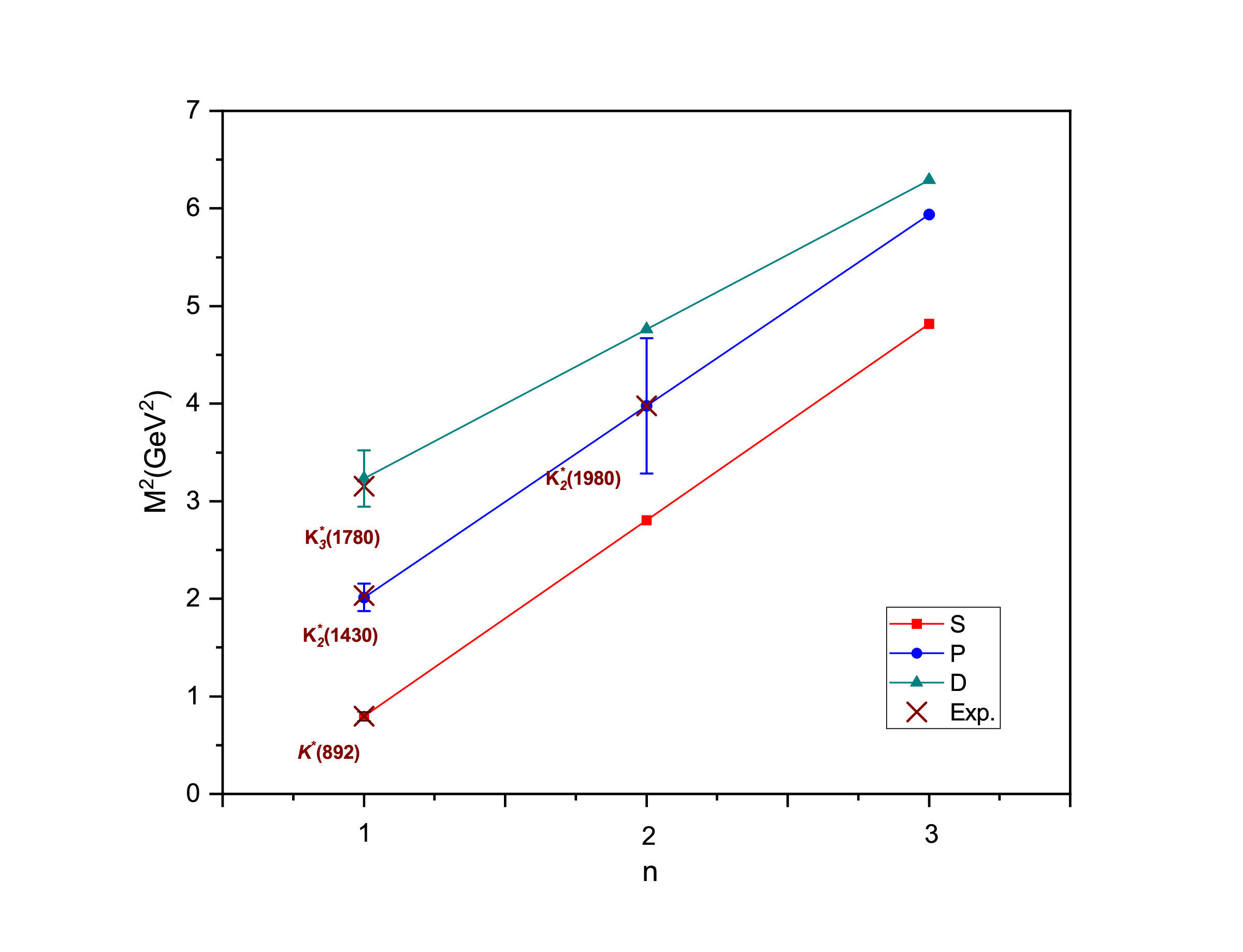}
	\caption{\label{fig:6}{Regge trajectories in the ($n,M^{2}$) plane for kaon with natural parity states. Error bars correspond to take $\Delta M^{2} = \pm \Gamma M$.}}
\end{figure*}
\begin{figure*}
	\centering
	\includegraphics[scale=0.3]{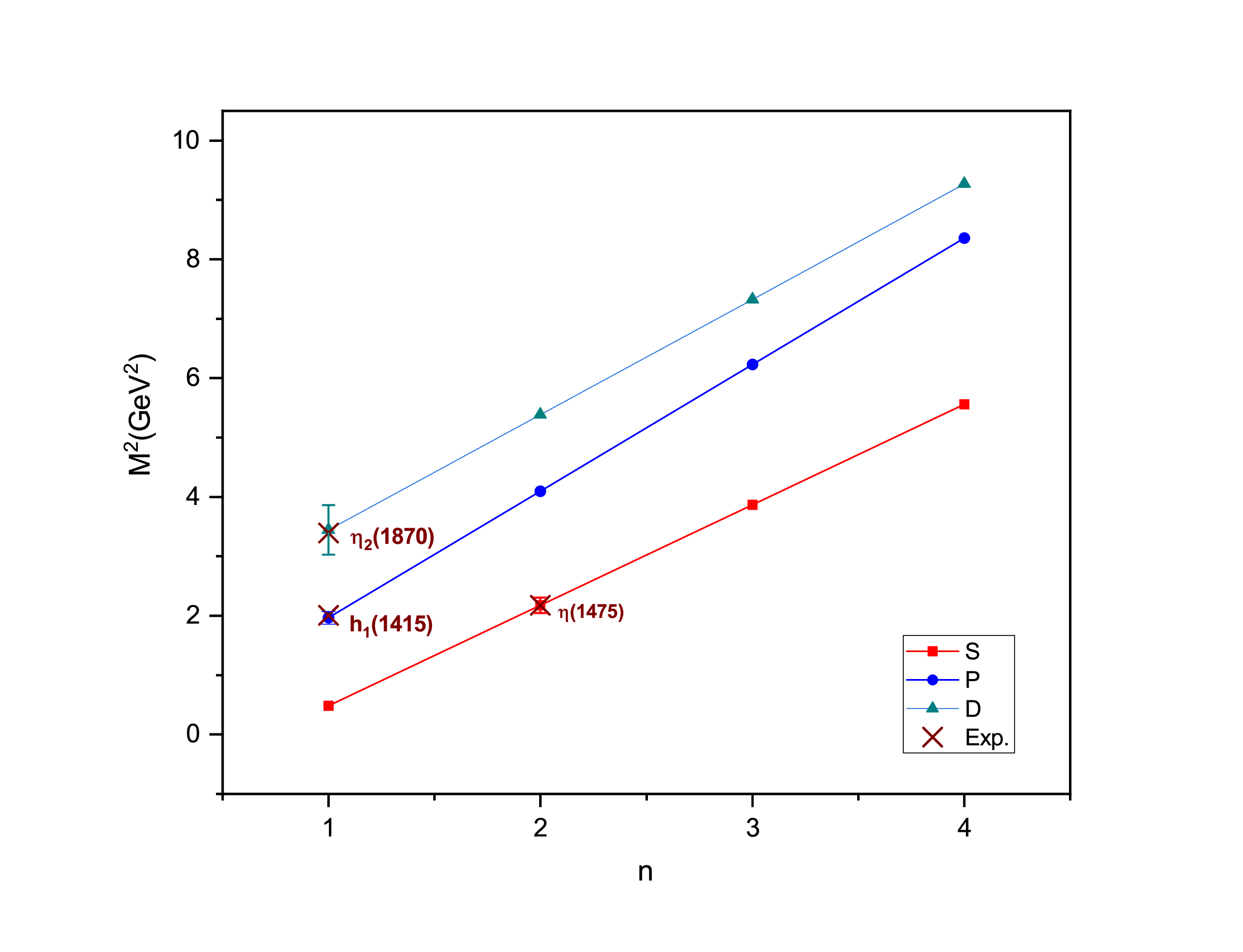}
	\caption{\label{fig:7}{Regge trajectories in the ($n,M^{2}$) plane for strangeonium with unnatural parity states. Error bars correspond to take $\Delta M^{2} = \pm \Gamma M$.}}
\end{figure*}
\begin{figure*}
	\centering
	\includegraphics[scale=0.3]{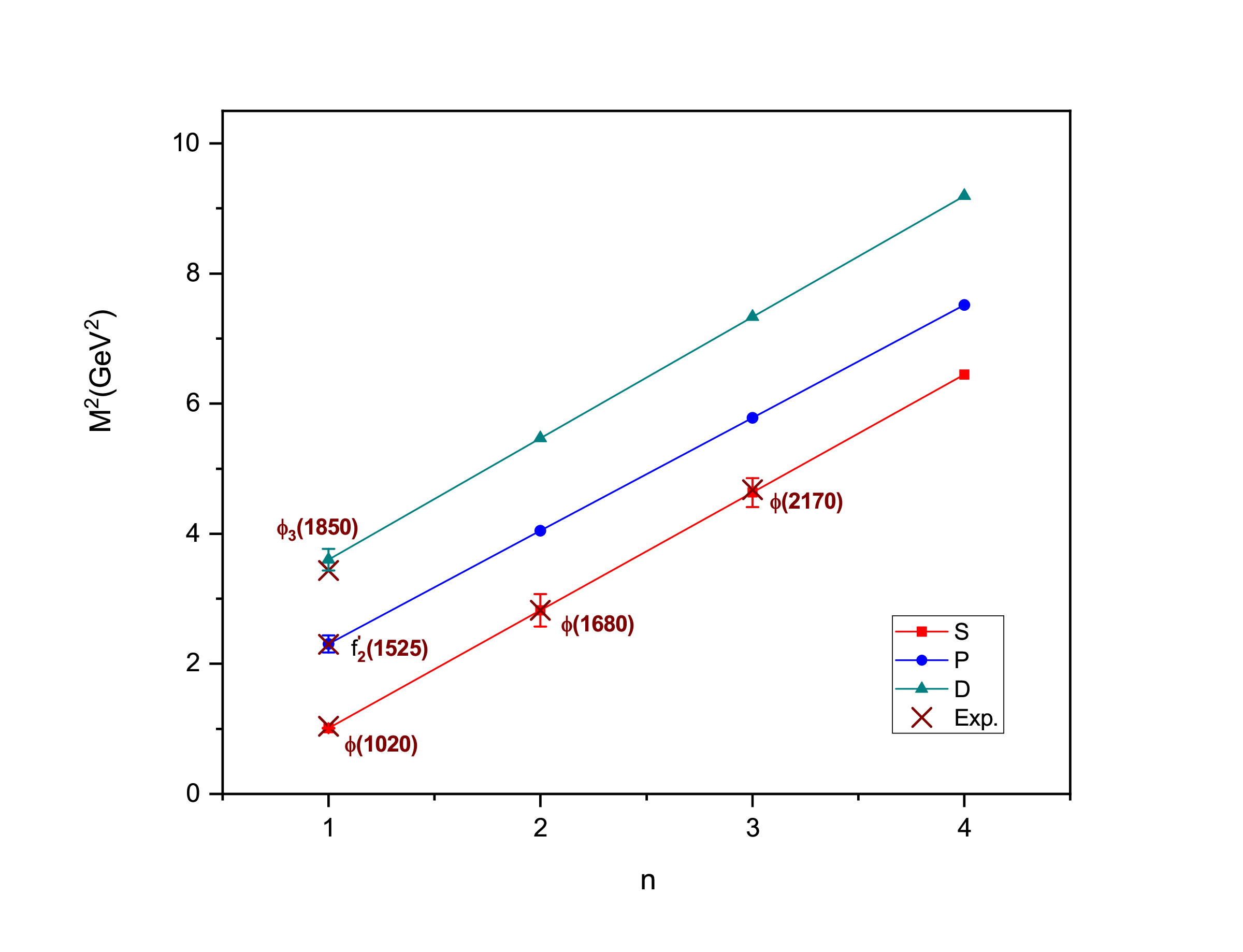}
	\caption{\label{fig:8}{Regge trajectories in the ($n,M^{2}$) plane for strangeonium with natural parity states. Error bars correspond to take $\Delta M^{2} = \pm \Gamma M$.}}
\end{figure*}

\section{Results and Discussion}
\begin{figure*}
	\hspace{-1cm}
	\includegraphics[scale=0.45]{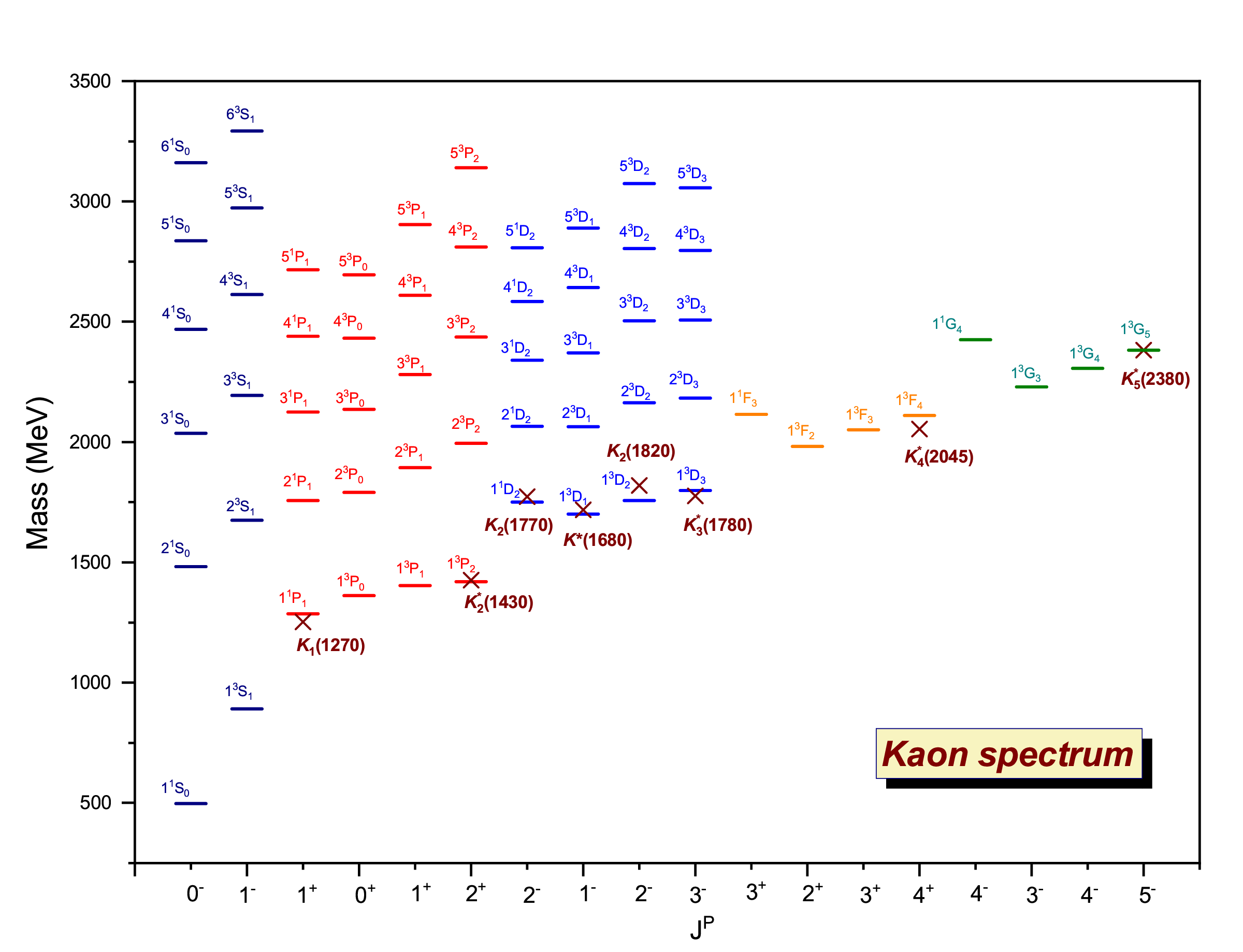}
	\caption{\label{fig:9}{The mass spectra of Kaon predicted by our model with comparison of experimentally established states mentioned in PDG \cite{PDG}. The coloured straight lines represents our predicted masses and the cross sign represents the experimental masses}}
\end{figure*}

The main purpose of the present work is to study the strange mesons (kaons) and the states which can be considered as pure $s\overline{s}$. Here we do not consider the states having strong flavour mixing between $n\overline{n} = (u\overline{u}+d\overline{d})/\sqrt{2}$ and $s\overline{s}$. 
The mass-spectra of these light flavoured mesons have been obtained successfully using Regge phenomenology. The numerical results are shown in Tables \ref{tab:table1} - \ref{tab:table4}. We have also given a global picture of the predicted kaon and strangeonium mass spectra along with the comparison of experimentally observed states for more clear picture, as shown in Figs. \ref{fig:9} and \ref{fig:10}. The colored straight lines represent the predicted masses obtained from our model, and the cross sign represents the experimentally detected masses mentioned in PDG \cite{PDG}. From Figs. \ref{fig:9} and \ref{fig:10} we can say that our estimated masses are very close to the experimental masses.  
The detailed discussion of the calculated theoretical predictions for kaons and strangeonium is given below.

\subsection{kaon spectrum}
With the amount of experimental data available for strange mesons, we compared our obtained results presented in tables \ref{tab:table1} and \ref{tab:table3} with well established  experimental masses and also with the predictions obtained from other theoretical approaches. The orbitally excited state masses calculated for kaons in the ($J,M^{2}$) plane are shown in table \ref{tab:table1}. The low lying $1P$ state $K_{1}(1270)$ having spin-parity $1^{+}$ listed in PDG \cite{PDG} with well established mass 1253$\pm$7 MeV is close to our predicted mass 1286.81$\pm$0.77 MeV for $1^{1}P_{1}$ state with a mass difference of 33.8 MeV. The studies \cite{Ebert2009,U. T. Nieto2022} also assigned $K_{1}(1270)$ resonance to be a $1^{1}P_{1}$ state. Also, our predicted mass shows a good agreement with the results obtained in the relativistic quark model \cite{Ebert2009} having a mass difference of only 8 MeV and slightly lower than the predictions of \cite{C.Q. Pang2017,U. T. Nieto2022,Godfrey1985}.

Another $1P$ state; $K_{2}^{*}(1430)$ having $J^{P}=2^{+}$ mentioned in PDG \cite{PDG} with experimental mass 1425.6$\pm$1.5 MeV which is assigned to be $1^{3}P_{2}$ by various theoretical studies \cite{Ebert2009,U. T. Nieto2022,Vijande12005} is observed to be very close to our calculated mass 1419.20$\pm$10.47 MeV for the $1^{3}P_{2}$ state having a mass difference of only 6.4 MeV. We compared our calculated mass value with those obtained by other theoretical models. Very less mass difference in the range of 5-12 MeV is seen with the predicted masses of Refs. \cite{Ebert2009,C.Q. Pang2017,Godfrey1985} and a slightly higher mass difference of around 50 MeV is shown with the results obtained in \cite{S. Ishida1987,Vijande12005}.

The strange meson $K(1680)$ reported in PDG \cite{PDG} having measured mass 1718$\pm$18 MeV with $J^{P}$ = $1^{-}$ can either belong to the $1^{3}D_{1}$ or $2^{3}S_{1}$ state. In the present work, the predicted  mass for $1^{3}D_{1}$ state as 1700.77$\pm$30.12 MeV agrees well with the experimental mass of $K(1680)$, hence we assign this resonance to the $1^{3}D_{1}$ state. 
The two experimentally well established $K$ meson resonances $K_{2}(1770)$ and $K_{2}^{*}(1820)$ listed in PDG with quantum numbers $J^{P} = 2^{-}$. Our mass value prediction for $1^{1}D_{2}$ state as 1750.46$\pm$0.80 MeV agrees very well with the experimental mass 1773$\pm$8 MeV with a mass difference of around 23 MeV. The predicted mass for $1^{3}D_{2}$ state is bit smaller than the experimental mass of $K_{2}^{*}(1820)$. Another candidate  $K_{3}^{*}(1780)$  reported in PDG having measured mass 1776$\pm$7 with sin-parity $3^{-}$ belongs to $1D$ state, and the model prediction for  $1^{3}D_{3}$ state as 1798.11$\pm$11.69 MeV is very close to the experimental mass having a mass difference of 22 MeV.
Also, our mass value predictions for $1D$-states shows a general agreement with the results obtained in other theoretical approaches \cite{Ebert2009,C.Q. Pang2017,U. T. Nieto2022,Godfrey1985,Vijande12005}.

The $K_{4}^{*}(2045)$ state collected in the PDG \cite{PDG} with the confirmed spin-parity $4^{+}$ which has mass 2054$\pm$9 MeV. Our obtained mass for $1^{3}F_{4}$ state is slightly higher than its experimentally observed mass having a mass difference of 56 MeV.  There is a agreement between our prediction by Regge phenomenology and the masses obtained by \cite{Ebert2009,U. T. Nieto2022,Godfrey1985}, while \cite{C.Q. Pang2017} predicts this state to be 52 MeV lower than our mass.
	The only state having spin-parity $5^{+}$ listed in PDG \cite{PDG} is $K_{5}^{*}(2380)$ with a  mass of 2383$\pm$24 MeV which was discovered in 1986 but has not been validated by other experimental collaborations and this resonance need more confirmation. Our calculated mass 2381.46$\pm$12.48 for $1^{3}G_{5}$ state is almost the same with the experimental mass of $K_{5}^{*}(2380)$ state having a mass difference of less than 1 MeV. Hence in this work, we predict this resonance belongs to $1^{3}G_{5}$ state. The results obtained in theoretical studies \cite{Ebert2009,Godfrey1985} well in agreement with our predicted mass having a mass difference of few MeV, whereas  the Refs. \cite{C.Q. Pang2017,U. T. Nieto2022} shows a mass difference of 90-95 MeV with our estimated mass. The remaining excited states whose experimental evidences are not yet seen show a general agreement with the results obtained in other theoretical approaches \cite{Ebert2009,C.Q. Pang2017,Godfrey1985}.
	
	Further, the radially excited state masses estimated in the present work for kaons are represented in the table \ref{tab:table3}. The obtained mass spectra for strange meson is compared with various theoretical approaches, and a general agreement can be seen with the results of Refs. \cite{C.Q. Pang2017,U. T. Nieto2022,Godfrey1985,Vijande12005}. 
	
Figs. \ref{fig:5} and \ref{fig:6} depicts the Regge plots for radial excitations. The trajectories for $S$, $P$, and $D$ states are not exactly parallel and equidistant. The uncertainty in the resonance masses are shown by the error bars and a quite large errors have been seen in the resonance masses of kaons.

	\subsection{Strangeonium spectrum}
	
	Since there are only few experimentally well established resonances which are acquired as pure $s\overline{s}$ states. The calculated ground state and the orbitally excited state masses in the ($J,M^{2}$) plane are summarized in the table \ref{tab:table2}. 
	The ground state ($1^{1}S_{0}$) mass 695.8 MeV predicted by our model is compared with  various results obtained in different theories. It ranges from 650-960 MeV. Some theories predict  $\eta^{'}$ state mentioned in PDG \cite{PDG} as the ground state mass of strangeonium \cite{Godfrey1985,Vijande12005}. But some of the recent studies denied to accept this state as a strangeonium, as it is a mixed flavour state \cite{Ebert2009,Q. Li2021,L. Y. Xiao2019,S. Ishida1987}. Our obtained mass is consistent with the prediction of \cite{Ebert2009,Q. Li2021,L. Y. Xiao2019,S. Ishida1987} having a mass difference of few MeV. The lowest $S$-wave vector state, $\phi$ which was first detected in a bubble chamber in 1962. It is well established and listed in PDG with $J^{P}=1^{-}$ having mass 1019.46$\pm$0.016 MeV. The estimated mass for $1^{3}S_{1}$ state as 1005.63 MeV is well produced by our employed theory and seems to be very close to the experimental mass of $\phi$ having a mass difference of 13.8 MeV only. Other than this our calculated mass is also in good agreement with the mass obtained for $1^{3}S_{1}$ state by various theoretical models \cite{Ebert2009,S. Ishida1987,Q. Li2021,L. Y. Xiao2019}. 
	The $h_{1}(1415)$ resonance listed by the PDG with $J^{P}=1^{+}$ and mass 1416$\pm$8 MeV. The recent BESIII measurements have improved the precision of the observed mass and width of $h_{1}(1415)$. Various theoretical studies predicted this state to be a good candidate of $1^{1}P_{1}$ $s\overline{s}$ state \cite{Ebert2009,Q. Li2021,L. Y. Xiao2019}. Our estimated mass 1402.01$\pm$0.78 MeV for $1^{1}P_{1}$ state is in well agreement with the experimental mass of $h_{1}(1415)$ having a mass difference of 14 MeV. While it shows slight disagreement with the  predictions of various theoretical models. Our estimated mass seems to be lower than the results obtained in \cite{Ebert2009,Vijande12005,Q. Li2021,L. Y. Xiao2019}.
	
	\begin{figure*}
			\hspace{-1cm}
		\includegraphics[scale=0.45]{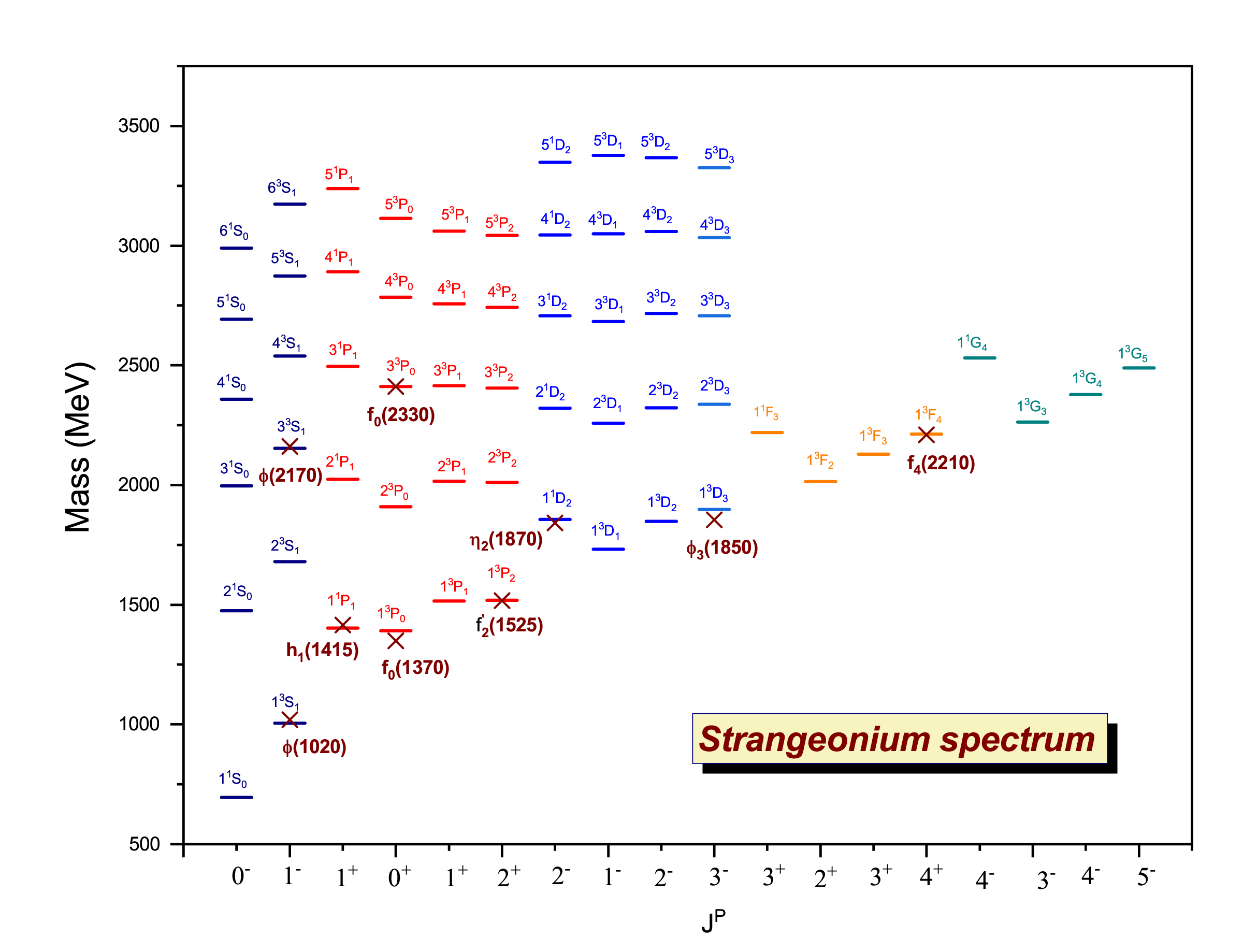}
		\caption{\label{fig:10}}{The mass spectra of strangeonium predicted by our model with comparison of experimentally established states mentioned in PDG \cite{PDG}. The coloured straight lines represents our predicted masses and the cross sign represents the experimental masses}
	\end{figure*}
	
	Another state $f_{2}^{'}(1525)$ which is widely accepted as the $1^{3}P_{2}$ $s\overline{s}$ state listed in the PDG \cite{PDG} with spin-parity $2^{+}$ having mass 1517.4$\pm$2.5 MeV is very close to the mass predicted by our model as 1518.70$\pm$17.67 MeV with a mass difference of only 1 MeV. Hence, the $J^{P}$ of the experimentally observed state $f_{2}^{'}(1525)$ is confirmed to be $2^{+}$ in this work. Also, our estimated mass is consistent and very close to the  outcomes of Refs. \cite{Ebert2009,Q. Li2021,L. Y. Xiao2019}.
	The experimental status of $1^{3}P_{0}$ $s\overline{s}$ state is a long-standing puzzle. The $0^{++}$ meson state $f_{0}(1370)$ mentioned in PDG \cite{PDG} has a mass in the range of $\approx$ 1200-1500 MeV. Recently, the BESIII Collaboration \cite{ssbar2} established $f_{0}(1370)$ state with the measured mass 1350$\pm$9$^{+12}_{-2}$ MeV which is close to our predicted mass 1392.0 MeV with a mass difference of 42 MeV. Hence, $f_{0}(1370)$ can be a good candidate for the $1^{3}P_{0}$ $s\overline{s}$ state. 
	
	The resonance $\eta_{2}(1870)$ listed in PDG \cite{PDG} with $J^{P}=2^{-}$, having averaged mass 1842$\pm$8 MeV is observed to be near to our estimated mass 1856.64$\pm$0.83 MeV for $1^{1}D_{2}$ state with a mass difference of nearly 15 MeV. The relativistic quark model theory also suggests $\eta_{2}(1870)$ resonance as $s\overline{s}$ state belongs to $1^{1}D_{2}$ state. Here, we assigned this state as $1D$ $s\overline{s}$ state with $J^{P}=2^{-}$ by our model.  Also, our obtained mass value prediction shows a general agreement with the masses generated by various theoretical approaches. The results in Refs. \cite{Vijande12005,Q. Li2021,S. Ishida1987} are close to the mass obtained by our model with a mass difference of few MeV, whereas the studies \cite{Ebert2009,L. Y. Xiao2019} shows a slight large mass difference.
	One more state $\phi_{3}(1850)$ which is accepted as $s\overline{s}$ resonance is assigned to be $1^{3}D_{3}$ state in recent theoretical studies \cite{Ebert2009,Q. Li2021}. This state was first identified in the $K\overline{K}$ invariant mass spectrum at CERN \cite{ssCERN} and further experiments at CERN \cite{ssSLac} and SLAC LASS \cite{ss3} determined its spin-parity to be $3^{-}$. Our estimated mass for $1^{3}D_{3}$ state as 1897.79$\pm$19.99 MeV is slightly higher than the experimentally observed mass of $\phi_{3}(1850)$ state as 1854$\pm$7 MeV mentioned in PDG \cite{PDG} with a mass difference of 43 MeV. The theoretical mass values  predicted for this state in the Refs. \cite{Vijande12005,L. Y. Xiao2019} are very close to the mass obtained by our model having a mass difference of few MeV,  while a slight large mass difference can be seen with the results obtained in studies \cite{Ebert2009,Q. Li2021,S. Ishida1987}. 
	
	The LASS Collaboration detected a narrow resonance $f_{4}(2210)$ with $J^{PC}$ value $4^{++}$ having a measured mass 2209$^{+17}_{-15}$ MeV \cite{LASS1988}. Later on, the mass and decay width of this resonance shows agreement with the results obtained at MARKIII Collaboration \cite{MARKIII} and WA67 (CERN SPS) \cite{PSLBooth1986}. Our predicted mass for $1^{3}F_{4}$ as 2212.87$\pm$21.01 MeV is very close to the experimentally detected mass \cite{LASS1988}. Hence, the resonance  $f_{4}(2210)$ might be a good candidate of $1^{3}F_{4}$ state. 
	
	Similarly, the radially excited state masses obtained in the present work for strangeonium are shown in table \ref{tab:table4}. We compared our estimated $s\overline{s}$ states with available experimentally observed masses and the mass spectra obtained by different theoretical models. Again a general trend of agreement is seen in case of $s\overline{s}$ states also. The $\phi(2170)$ resonance which is mentioned in PDG \cite{PDG} with spin-parity $1^{-}$, is a possible radial $s\overline{s}$ state. Our calculated mass for $3^{3}S_{1}$ state as 2152.56$\pm$39.98 MeV is very close to the its experimental mass 2162$\pm$7 MeV having a mass difference of nearly 10 MeV. Hence we assigned this state to be a good candidate of $3^{3}S_{1}$ state. In the relativistic quark model \cite{Ebert2009}, the authors also suggests this state as a radial state belongs to $3^{3}S_{1}$. Also, one more state $f_{0}(2330)$ listed in PDG with $J^{PC}$ = $0^{++}$ needs more confirmation and the mass of this state is still not established. Our predicted mass for $3^{3}P_{0}$ state as 2411.69 MeV matches exactly same with the mass observed at BESIII which is 2411$\pm$10$\pm$7 MeV by an amplitude analysis of the $K_{S}K_{S}$ system produced in radiative $J/\psi$ decays \cite{ssbar2}.  
	Also, our obtained mass is very close to the recent observation of \cite{Ricken2003} which predicts the mass of $f_{0}(2330)$ as 2419$\pm$64 MeV. Hence the resonance $f_{0}(2330)$ listed by PDG may belongs to $3^{3}P_{0}$ $s\overline{s}$ state.
	Also, the predicted mass spectra shows good agreement with the outcomes of Refs. \cite{L. Y. Xiao2019,S. Ishida1987} with mass difference of few MeV, while some discrepancies can be shown with the results of Refs. \cite{Godfrey1985,Vijande12005}.
	
The Regge trajectories plotted for the strangeonium meson in the ($n,M^{2}$) plane are shown in Figs. \ref{fig:7} and \ref{fig:8}. A slight deviation is observed in the trajectories drawn for $S$, $P$, and $D$ states from being exactly parallel and equally distant. The error bars in the line shows uncertainty in the masses.
	

\section{Conclusion}
Our model have been successfully employed to evaluate the spectrum of kaon and strangeonium. Unlike strangeonium, kaon have a large number of experimentally established states. In the present work we confirmed the spin-parity of $K_{1}(1270)$, $K_{2}(1770)$, $K_{2}^{*}(1430)$, $K^{*}(1680)$, $K_{2}(1820)$, $K_{3}^{*}(1780)$ and $K_{4}^{*}(2045)$, which also gives a validation to the evaluated results from our employed model. The $K_{5}^{*}(2380)$ state needs more confirmation and we predict the quantum numbers of this resonance which might be useful for future experimental progress. 

Since, only few experimental data is available for pure strangeonium states and the $J^{P}$ values of $\phi$, $h_{1}(1415)$, $\eta_{2}(1870)$, $f_{2}^{'}(1525)$, and $\phi_{3}(1850)$ states is confirmed in this work.
Some recently observed resonances by various experimental groups still needs further verification and yet to be confirmed in near future. In the present work we try to predict the possible quantum numbers of such  states which might be the candidates of strangeonium states namely; $f_{0}(1370)$, $\phi(2170)$, $f_{0}(2330)$, $f_{4}(2210)$. might be the candidates of $s\overline{s}$ states, which will provide a valuable contribution in future experiments.

The reliability of the model depends upon the mass inputs we have taken for calculation. When the experimental masses are taken as inputs the estimated results are very close to the experimentally observed masses. But when we taken the theoretical masses as inputs our calculated shows bit discrepancies with the experimental masses. The non-parallelism in the plots of $n$ and $M^{2}$ is clearly observed. We have incorporated the half width rule as a uncertainty in the resonance masses. A quite large error bars are observed in the Regge lines of kaons which are more deviated as compared to the trajectories of strangeonium mesons where deviation is less and uncertainty in the masses are small. 

Figs. \ref{fig:9} and \ref{fig:10} clearly depicts that our obtained results are very close to the experimental observations where available. Also, the evaluated  mass-spectra shows a general agreement with the predictions of other theoretical approaches. Hence, this study with large number of mass predictions might be helpful and will provides important clues to the experimental facilities such as BESIII, LHCb, BABAR, JPARC etc. in future searches and the missing excited states.

	
	%
	%
	\appendix
	
	%


\end{document}